\documentclass[12pt,preprint]{aastex}

\def\dw#1{}			
\def\jm#1{}

\newcommand{\op}{Ly$\alpha$\ }

\begin{document}


\title{Small Scale Structure at High Redshift:
II. Physical Properties of the C~IV Absorbing Clouds\altaffilmark{1}}


\author{Michael Rauch\altaffilmark{2}, Wallace L.W. Sargent\altaffilmark{3},
Tom A. Barlow\altaffilmark{3}
}
\altaffiltext{1}{The observations were made at the W.M. Keck Observatory
which is operated as a scientific partnership between the California
Institute of Technology and the University of California; it was made
possible by the generous support of the W.M. Keck Foundation.}
\altaffiltext{2}{Carnegie Observatories, 813 Santa Barbara Street,
Pasadena, CA 91101, USA}
\altaffiltext{3}{Astronomy Department, California Institute of Technology,
Pasadena, CA 91125, USA}

\medskip 
\vskip 2.5cm



\vfill

\pagebreak

\begin{abstract}
Keck HIRES spectra with a resolution of 6.6 or 4.3 km s$^{-1}$ were 
obtained of the separate images of three gravitationally lensed QSOs,
namely Q0142-0959 A,B (UM 763, A,B) ($z_{em} = 2.72$), Q1104-1804 A,B
($z_{em} = 2.32$), and Q1422+2309 A,C ($z_{em} = 3.62$). The typical 
separation of the images on the sky is $\sim 1{\arcsec}$. The 
corresponding transverse distances between the lines of sight range
from a few tens of kpc at the redshift of the lens to a few pc at the source.
We studied the velocity differences and column density differences in
C~IV doublets in each QSO, including single, isolated doublets, complex
clumps of doublets and sub-clumps. Unlike the low ionization gas clouds 
typical of the
interstellar gas in the Galaxy or damped Ly$\alpha$ galaxies, the spatial 
density
distribution of CIV absorbing gas clouds turns out to be mostly
featureless on scales up to  a few hundred parsecs, with column density
differences rising to 50\% or more over separations beyond a few kpc.
Similarly, velocity shear becomes detectable only over distances larger
than a few hundred pc, rising to $\sim 70$ kms$^{-1}$ at a few kpc.
The absorption systems become more coherent with decreasing redshift
distance to the background QSO; this finding confirms that all 3 QSOs
used are indeed lensed, as opposed to being genuine QSO pairs.

The amount of turbulence in CIV gas along and across each line of sight
was measured, and a crude estimate of the energy input rate obtained.
The energy transmitted to the gas is substantially less than in present
day star-forming regions, and the gas is less turbulent on a given
spatial scale than, e.g., local HII regions.  The quiescence of CIV
clouds, taken with their probable low density, imply that these objects
are not internal to galaxies. The C~IV absorbers could be gas expelled
recently to large radii and raining back onto its parent galaxy, or
pre-enriched gas from an earlier (population III ?) episode of star
formation, falling into the nearest mass concentration. However, while
the metals in the gas may have been formed at higher redshifts ($z >
5$?), the residual turbulence in the clouds and the minimum coherence
length measured here imply that the gas was stirred more recently,
possibly by star formation events recurring on a timescale on the order
of $10^7- 10^8$ years.

\end{abstract}


\pagebreak

\section{Introduction}

In the second paper of a series on small scale structure in the baryon
distribution at high redshift, we study metal-enriched gas as seen in
absorption by CIV against multiple images of gravitationally lensed
QSOs. Paper I (Rauch, Sargent and Barlow 1999) dealt with a particular
low ionization metal system at redshift 3.58, which showed density and
velocity structure on scales of a few tens of parsecs, consistent with
a gas outflow on a very small (stellar) scale, possibly a supernova
remnant. The density, metallicity, and kinematics of the absorption
system resembled the ISM in our galaxy.  However, such strong, low
ionization (ISM) regions of interstellar gas are relatively rarely
traversed by random QSO lines of sight. High ionization CIV absorption
systems outnumber them by about 2 orders of magnitude (down to column
density limits of N(CIV)$\sim$ a few$\times$ 10$^{12}$cm$^{-2}$).  In
the present paper we turn our attention to the statistical properties
of the high ionization CIV gas phase.

The CIV $(\lambda \lambda$ 1549,1550 \AA ) doublet in absorption is the
most commonly observed signature of metal enriched gas at high
redshift. Most of the CIV systems seen belong to highly ionized gas,
optically thin at the HI Lyman Limit. Line widths are consistent with a
photoionized gas ($T \sim$ a few $\times 10^4$ K) with about equal
contributions from thermal and non-thermal motion (Rauch et al. 1996). At redshift z =
3, densities of gas with detectable CIV absoprtion appear to range
between a few times the mean density of the universe to values
somewhat below those of a virialized galactic halo (Rauch, Haehnelt \&
Steinmetz 1997; assuming photoionization by a standard QSO UV
background).  The CIV absorption components are highly clustered on
scales of several hundred kms$^{-1}$ (Sargent et al 1979; Steidel 1990;
Petitjean \& Bergeron 1994).

The stronger CIV absorption systems studied by several surveys during
the past two decades have often been interpreted as gas clouds orbiting
in hypothetical, gaseous galactic halos (e.g., Sargent et al.  1979;
Bergeron \& Boiss\'e 1991; Steidel 1993; Petitjean \& Bergeron 1994; Mo
\& Miralda-Escud\'e 1996).  The physical origin of the CIV clouds
themselves and their relation (if any) to known present-day Galactic
environments is not clear. They may conceivably be related to
proto-globular clusters (e.g, Fall \& Rees 1985; Kang et al 1990; Peng
\& Weisheit 1991), but it is also possible that they are caused by
galactic winds  (e.g., Fransson \& Epstein 1982; Wang 1995), or tidal
debris (Wang 1993; Morris \& Van den Bergh 1994).

Recent work with the Keck HIRES instrument (Cowie et al. 1995; Tytler
et al.  1995; Womble et al 1996; Cowie \& Songaila 1998) has detected
CIV absorption even in the diffuse, low density gas clouds giving rise
to the Ly$\alpha$ forest. These structures turned out to  already being
metal-polluted at high redshift (at the 50\% level for gas with a neutral hydrogen
column density $> 10^{14.5}$ cm$^{-2}$), reaching a median metallicity
$Z\sim 10^{-2.5} Z_{\odot}$ by redshift $z\sim 3$, albeit with a substantial scatter of about
a magnitude.  There is evidence for some metal enrichment at even lower
column densities (Ellison et al 2000, Schaye et al 2000).  The high
rate of incidence together with the large velocity widths and inferred
low gas densities of many CIV systems argue for a more complex origin
of the absorbers than the simple picture of gas clouds in a
hydrostatically supported halo can provide.  Continued infall or
outflows of metal enriched gas beyond the virial radius of high
redshift galaxies may be important.

A CDM -based scenario where CIV absorption arises in metal-enriched gas
accreting onto and shocked by forming galaxies has been discussed by
Rauch, Haehnelt, und Steinmetz (1997). In this model CIV systems at
high redshift correspond to groups of protogalactic clumps prior
to merging, which are the building blocks of future normal galaxies. In
this interpretation, the large velocity widths of CIV absorption
systems are a natural consequence of mergers between two or more clumps
or chance alignments of filamentary groups of such objects along our
line of sight.  The basic limitation of the original version of this
model is its reliance on gravitational instability as the sole
mechanism for structure formation.  Stellar 'feedback' (in the form of
discrete sources of ionization, momentum and energy input, and stellar
metal production) is capable of strongly modifying the  gas
distribution locally, e.g., it may produce outflows locally superseding
the general infalling motion.  We expect the observed
properties of the CIV gas-- its ionization state, kinematics and small
scale texture-- to reveal its relation to the stellar
outflows on one hand and the intergalactic baryonic reservoir on the
other. In particular we would like to determine the density fluctuations
and characteristic
sizes of the high ionization (CIV) gas clouds and the magnitude of the
internal velocity gradients as measured in projection. As in paper I we
employ observations of absorption lines in spectra of the separate
images of gravitationally lensed QSOs to obtain these constraints.

Earlier lensing studies have produced important constraints on the size
scales of the intergalactic gas clouds giving rise to the Ly$\alpha$
forest. Observations of lensed QSOs (Foltz et al.  1983; Smette et al.
1993,1995), taken together with similar constraints from the
observation of QSO pairs at larger separations (e.g., Sargent et al.
1982; Shaver \& Robertson 1983; Bechtold et al. 1994; Dinshaw et al.
1994,1995; Fang et al. 1996, Petry et al 1998)  indicate the presence
of large gaseous bodies, coherent over several hundred kiloparsecs.  In
contrast, the objects responsible for metal absorption systems
(especially damped \op systems) appear to have sizes up to a few tens
of kpc and smaller differences in their equivalent widths across the
lines of sight (Young et al. 1981; Smette et al. 1993,1995; Crotts et
al. 1994; Bechtold \& Yee 1995; Zuo et al. 1997; Michalitsianos et al.
1997; Monier, Turnshek \& Lupie 1998; Lopez et al 1997; Petitjean et al
2000). 

Before the advent of the Keck telescope observations of absorption in
multiple lines of sight at high redshift have
been very difficult and were limited to a resolution too low to make
strong inferences about the structure of metal systems. Using the Keck
High REsolution Spectrograph (HIRES; Vogt et al. 1994) we have begun a
survey of all gravitationally lensed QSOs bright enough to be observed
with HIRES ($m < 19$) and visible from Hawaii, which have image
separations wide enough to be separated in good seeing ($> 1"$). In
this paper we present the results of our observations of three QSOs--
Q0142-0959 A,B (UM 763, A,B) ($z_{em} = 2.72$), Q1104-1804 A,B ($z_{em}
= 2.32$), and Q1422+2309 A,C ($z_{em} = 3.62$). A convenient summary of
the observed properties of each QSO may be found on the CASTLES web
site (http://cfa-www.harvard.edu/castles/ ). Details of the
observations and data analysis are given in section 2, the results are
discussed in section 3 and the conclusions are summarised in section
4.

\section{Observations and Data Analysis}

The three brightest QSOs in our sample, with redshifts ($z>1.6$)
suitable for for the study of CIV absorption are UM673 (Surdej et al
1988), HE104-1805 (Wisotzki et al 1993), and Q1422+2309 (Patnaik et al 1992).
The spectra cover the wavelength ranges $\lambda\lambda$3608-6094
(UM 273 and HE1104-1805) and $\lambda\lambda$4832-7348 (Q1422+2309).
For UM673 and HE1104-1805 we used a 7"$\times 0.86"$ decker, giving a
spectral resolution of 6.6 kms$^{-1}$ (FWHM).  For Q1422+231, because
of the small separation of the images (1.3"), a narrower slit
(7"$\times 0.574"$) was used to reduce mutual contamination of the
spectra.  All spectra were taken with the red crossdisperser, which was
the only one available when the observations started (1996). The
individual images were put separately on the slit to facilitate
sky-background subtraction and offslit guiding. Guiding was done
mostly by hand because the automatic guider was found to be 
unreliable for this particularly demanding type of observation. The 
slit was aligned as closely as possible to the parallactic angle,
consistent with minimizing the spillover from the image not being
observed.

The A (B) image of UM673 was exposed for 13500s (33100s), the A (B)
image of HE1104-1805 for 19300s (51200s), and the A (C) image of
Q1422+2309 for 19600s (33200s), respectively. The A and B images of
Q1422+2309 were judged to be too close together for reliable 
observation of the B image.
We note that the main observational difficulty is to obtain satisfactory
spectra of the fainter image of each pair. The extra exposure times
listed above for the fainter images do not completely compensate
for the differences in brightness of the two images.

The data were reduced using the HIRES data reduction package MAKEE
written by one of us (TAB; see Barlow \& Sargent 1996).  For each QSO
the spectra from the different echelle orders of each image were
combined. The continua redward of the Ly$\alpha$ emission line of both
images were normalized by using polynomial fits to spectral regions
apparently free  of absorption lines. 

Two examples of strong CIV systems toward UM673 are shown in Fig.
\ref{CIV_UM673_velspec}. The systems are at redshifts $z_{\mathrm abs} = 1.94$
(top; separation between the lines of sight: 1.7 $h^{-1}$ kpc) and
2.355 (bottom; separation between the lines of sight: 0.63 $h^{-1}$
kpc). Both objects show remarkably large velocity extents (510 and 260
kms$^{-1}$) along the line of sight. In the upper image the first group
of absorption lines ($0<v<250$ kms$^{-1}$ on the arbitrary velocity
scale) appears to be affected by a coherent velocity shear across the
lines of sight. In contrast, the sharp, 'single' absorption doublet
near 450 kms$^{-1}$ is at nearly the same position in both images. Such
an absorption pattern could conceivably arise during the merger of two
galaxies which happen to be moving approximately along our line of
sight. The large velocity extent along the line of sight may be caused
by the infalling motion, and rotation in the first object may be
causing the velocity shear in the lower redshift clump.  In the bottom
spectrum ($z_{\mathrm abs} = 2.355$), the differences are somewhat less
dramatic, and may possibly be explained by fluctuations in the gas
density internal to a single galaxy.

To quantify line of sight differences in the CIV absorption systems we
proceeded as follows:  For the further analysis, only the regions of
the spectra longward of Ly$\alpha$ emission were used. CIV systems were
identified.  It is hard to estimate an overall completeness limit,
since the S/N ratio varies along the spectra, and Q1422+231 had a
significantly higher S/N ratio than the other two QSOs. Trials indicate
that any (unblended) CIV doublet  with a Doppler parameter of order 7
kms$^{-1}$ and with N(CIV)$\geq 5\times 10^{12}$ cm$^{-2}$ would have
been seen in most of the spectra of the three objects.

\begin{figure}
     \centering
  \includegraphics[width=5.5cm, angle=-90]{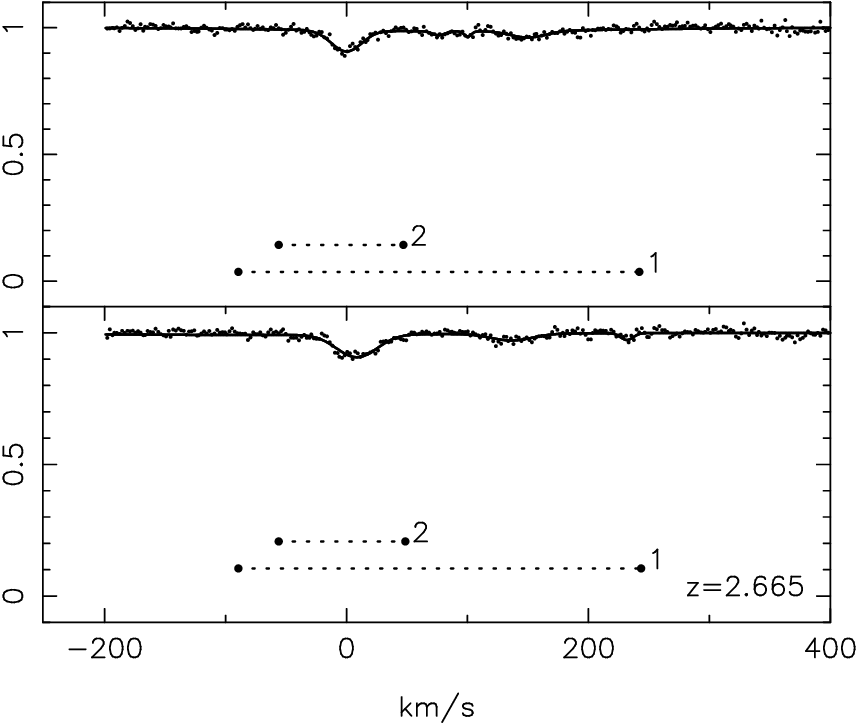}
  \includegraphics[width=5.5cm,angle=-90]{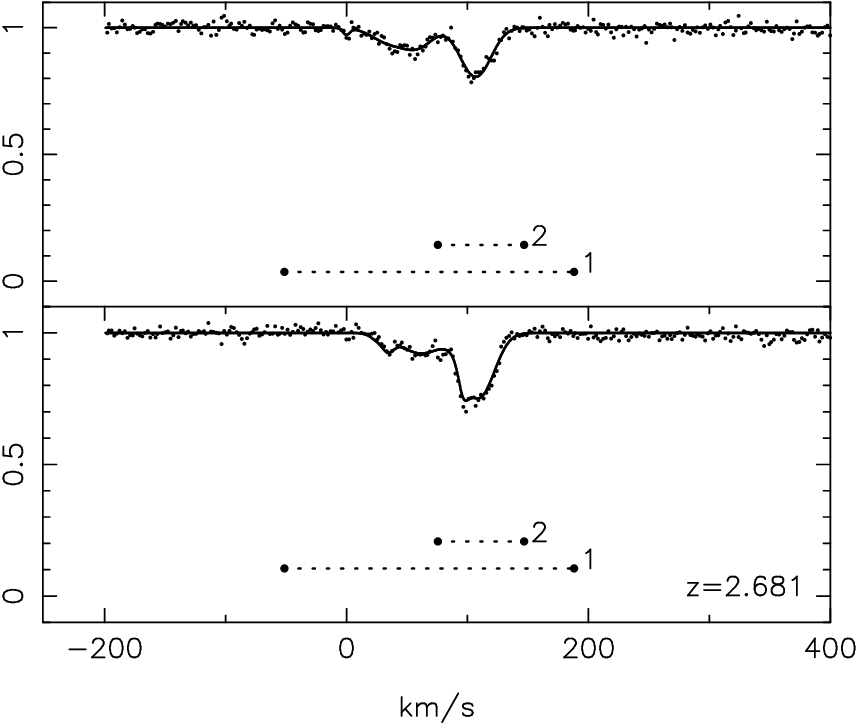}
  \includegraphics[width=5.5cm, angle=-90]{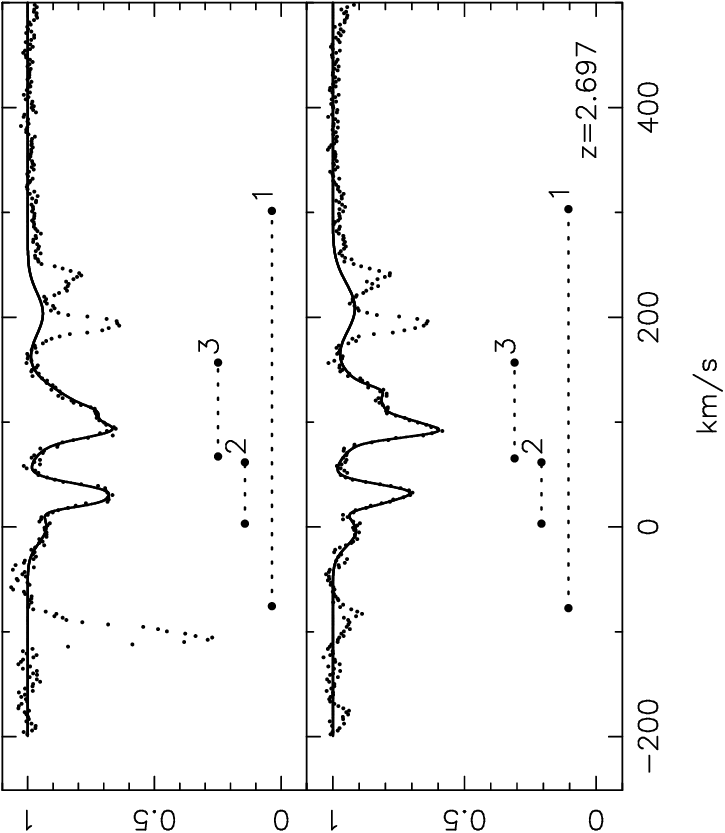}
  \includegraphics[width=5.5cm,angle=-90]{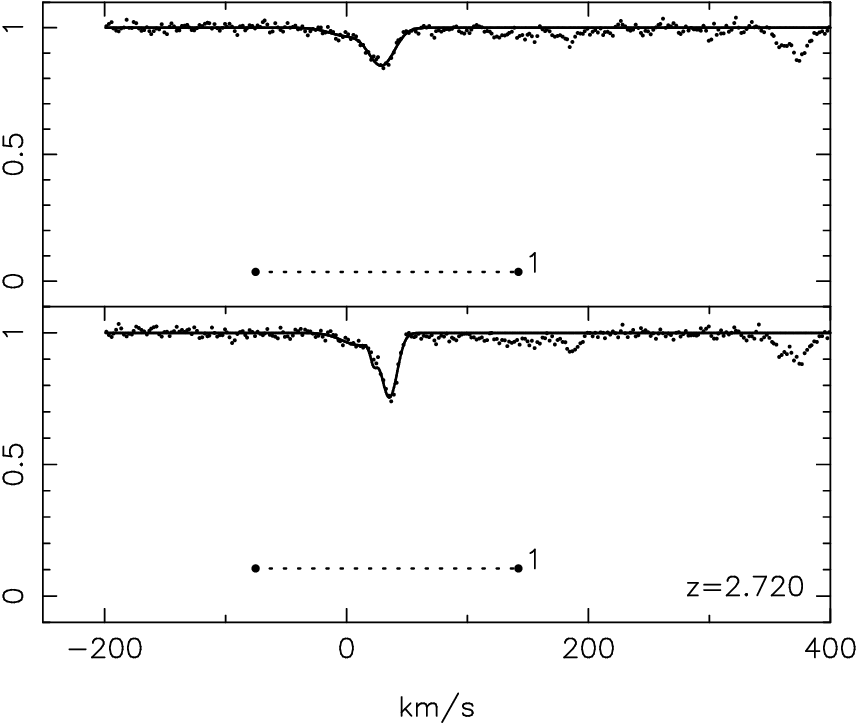}
  \includegraphics[width=5.5cm, angle=-90]{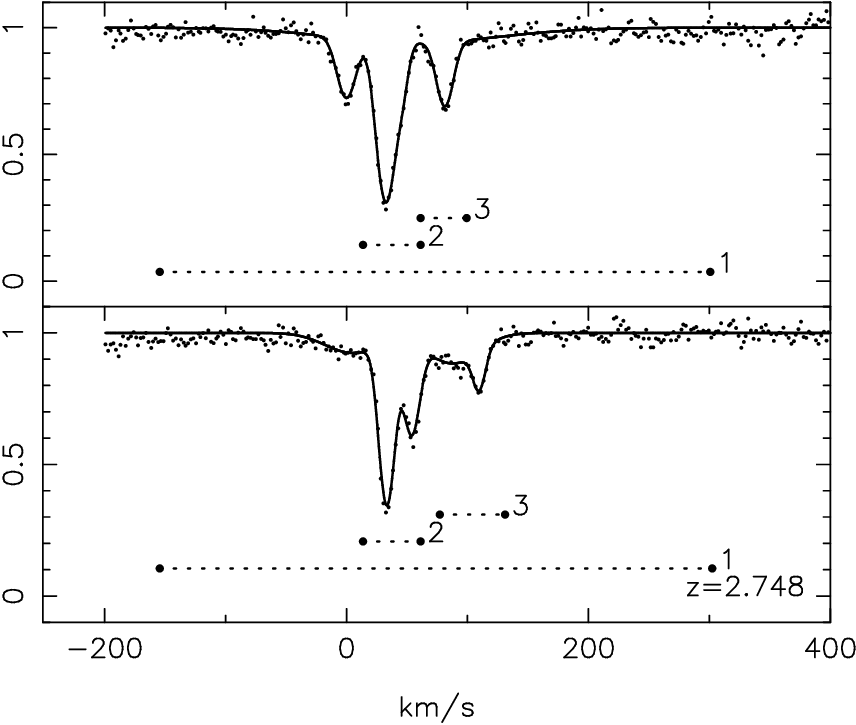}
  \includegraphics[width=5.5cm,angle=-90]{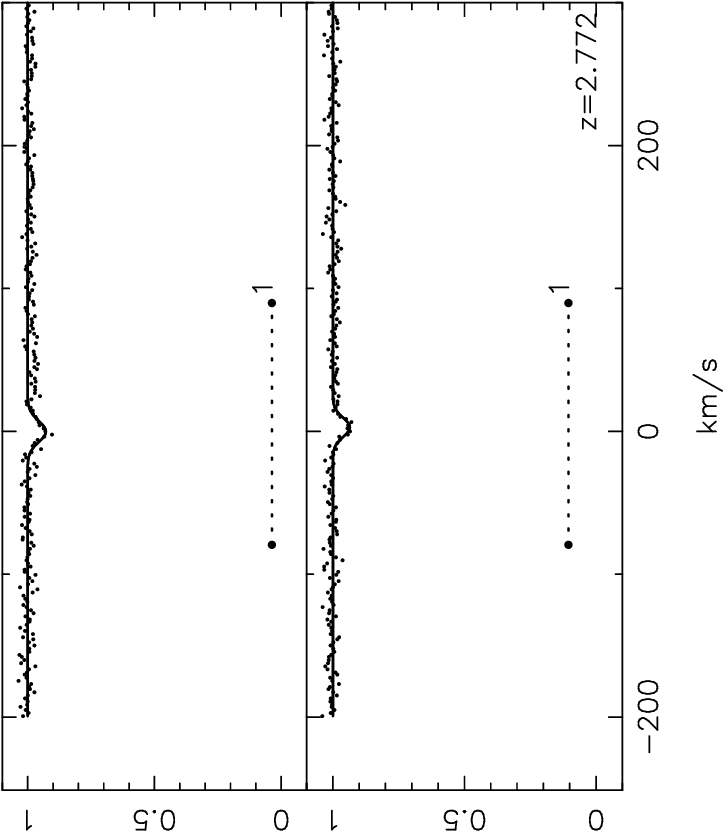}
     \caption{\small
CIV systems redward of the \op emission line in the A and C lines of
sight towards Q1422+231. In figures 2 -- 7, the solid line in the top and bottom plots
represents the Voigt profile models of the CIV $\lambda 1548
\AA\ $ transition in the two lines of sight. The models are plotted on
top of the actual data (thin dotted line).  The redshift is given in
the bottom RHS corner.  The horizontal dotted lines drawn underneath
the spectra mark the extent of the velocity windows in which the
integrated column densities and column density weighted velocities
(listed in tables 1 -- 3) were measured. The numbers correspond to
those in the first column of these tables, for each redshift. Any absorption
features not shown as fits are interlopers from other metal transitions.  All blends
between CIV and such lines have been fitted simultaneously but the results
are shown only for CIV to avoid confusion.  \label{CIV_1422}}
\end{figure}

\begin{figure}
     \centering
  \includegraphics[width=5.5cm, angle=-90]{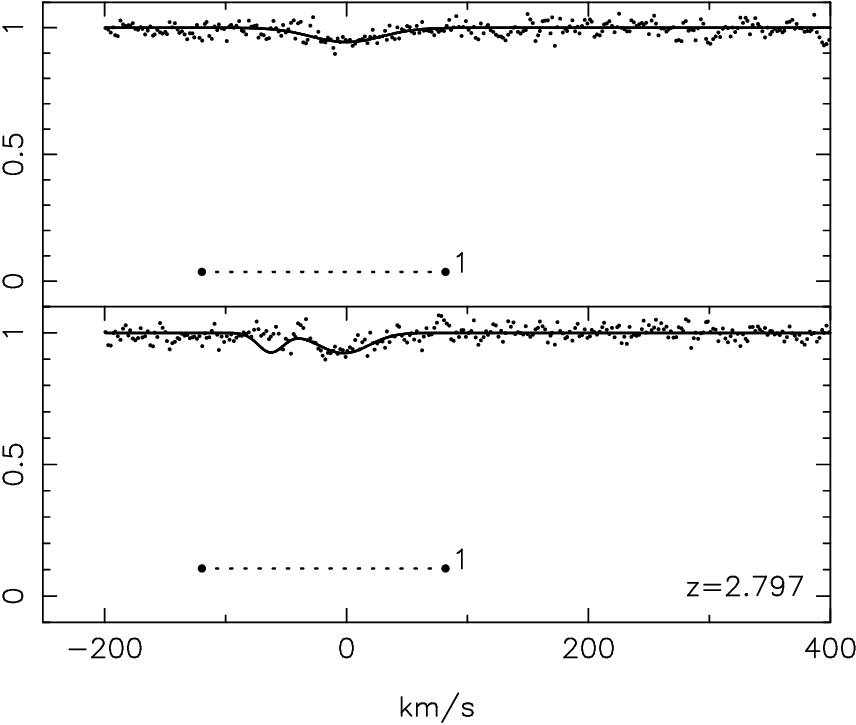}
  \includegraphics[width=5.5cm,angle=-90]{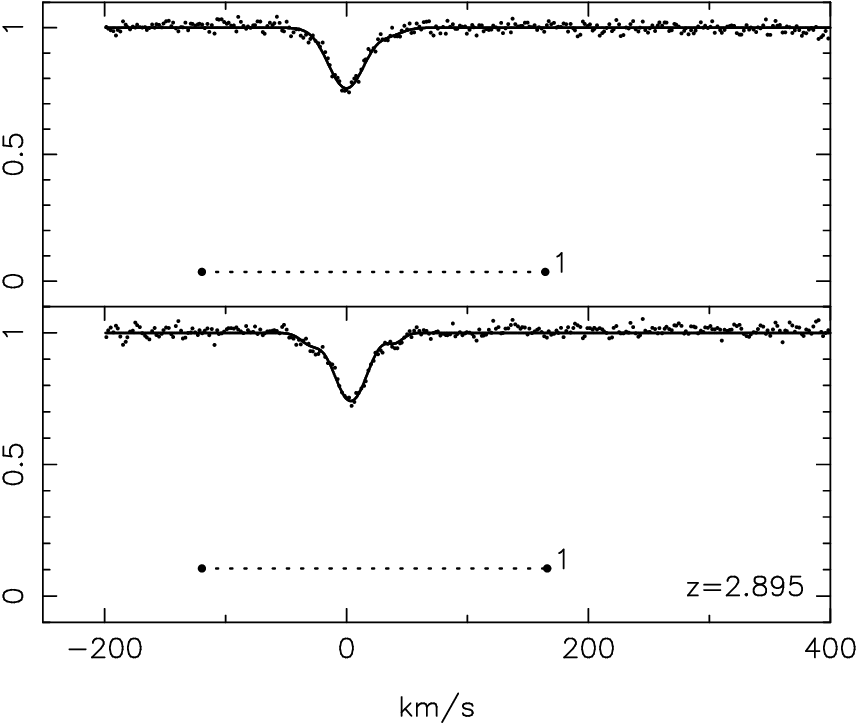}
  \includegraphics[width=5.5cm, angle=-90]{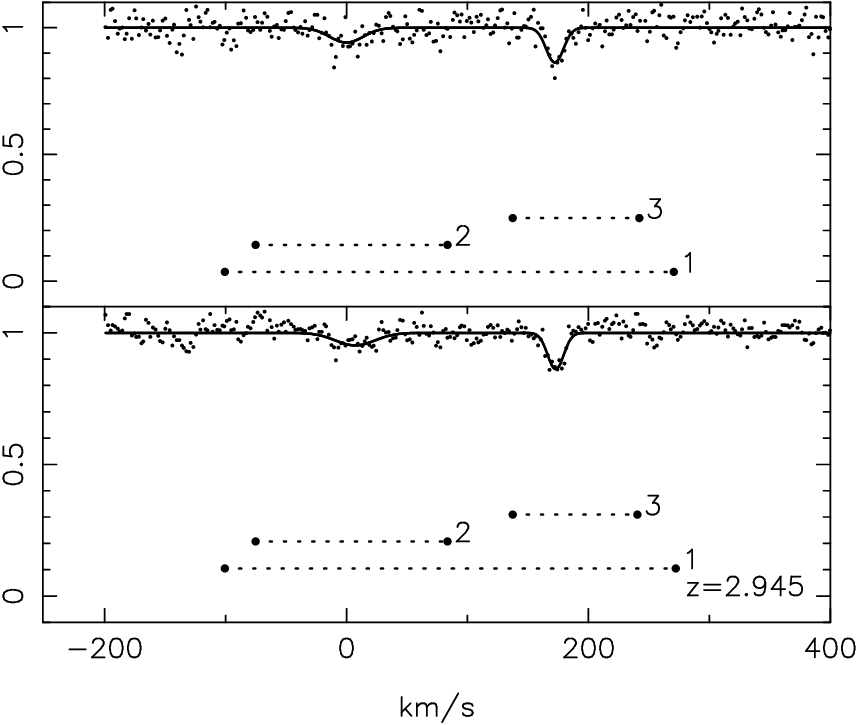}
  \includegraphics[width=5.5cm,angle=-90]{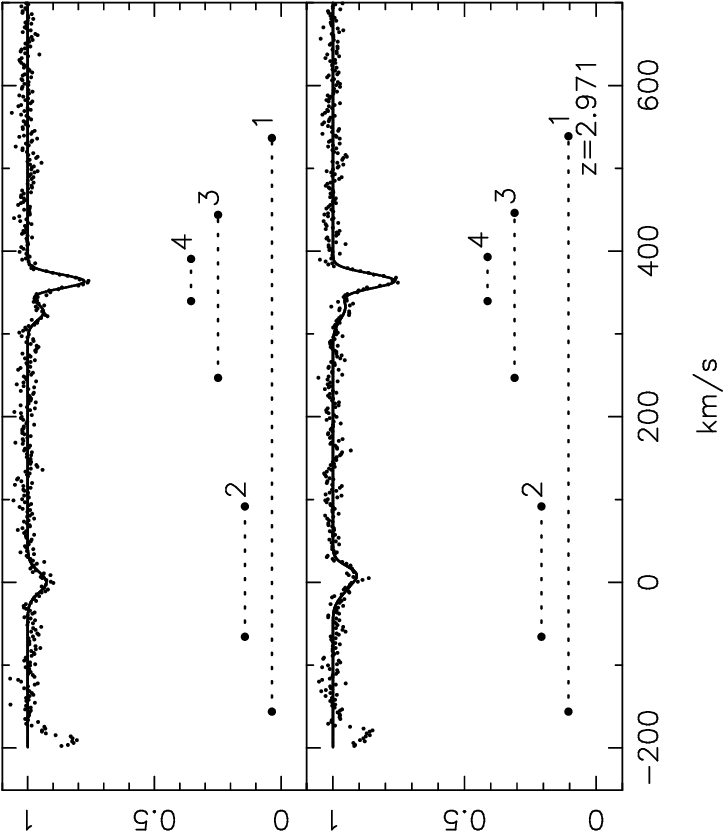}
  \includegraphics[width=5.5cm, angle=-90]{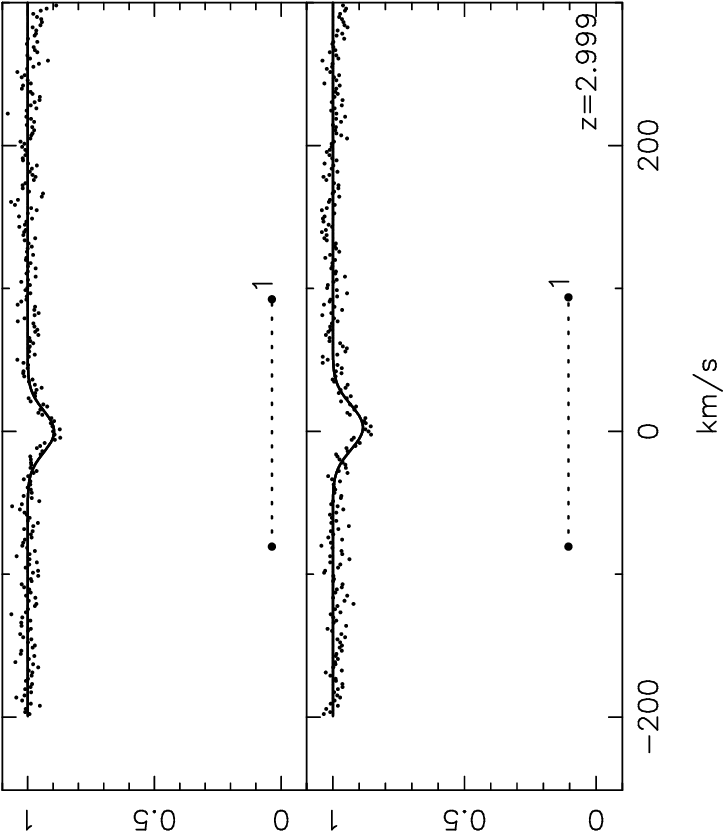}
  \includegraphics[width=5.5cm,angle=-90]{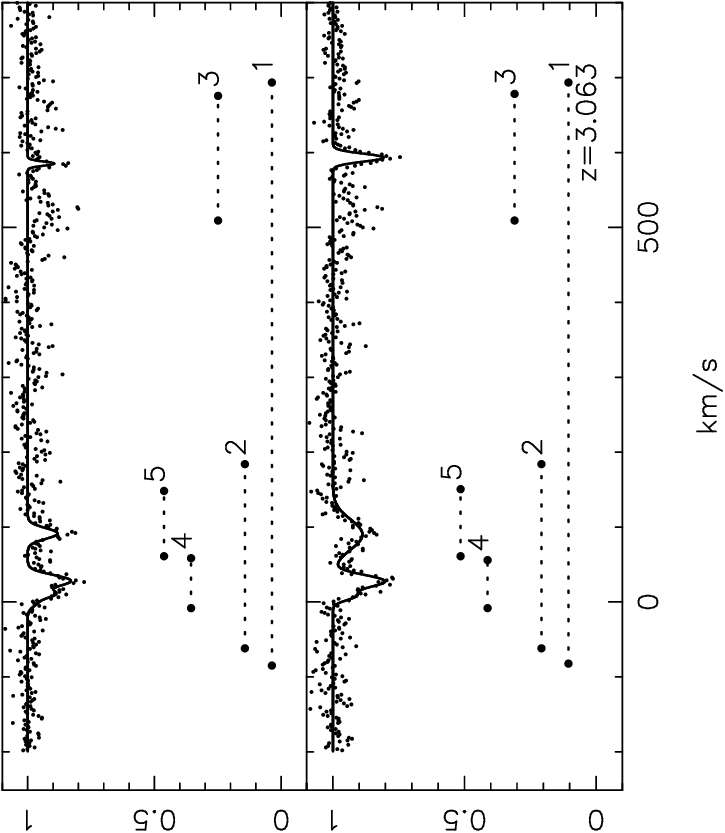}
     \caption{\small CIV systems redward of the \op emission line in the A and C lines of sight
towards Q1422+231 (continued)
\label{CIV_1422a}}
\end{figure}

\begin{figure}
     \centering
  \includegraphics[width=5.5cm, angle=-90]{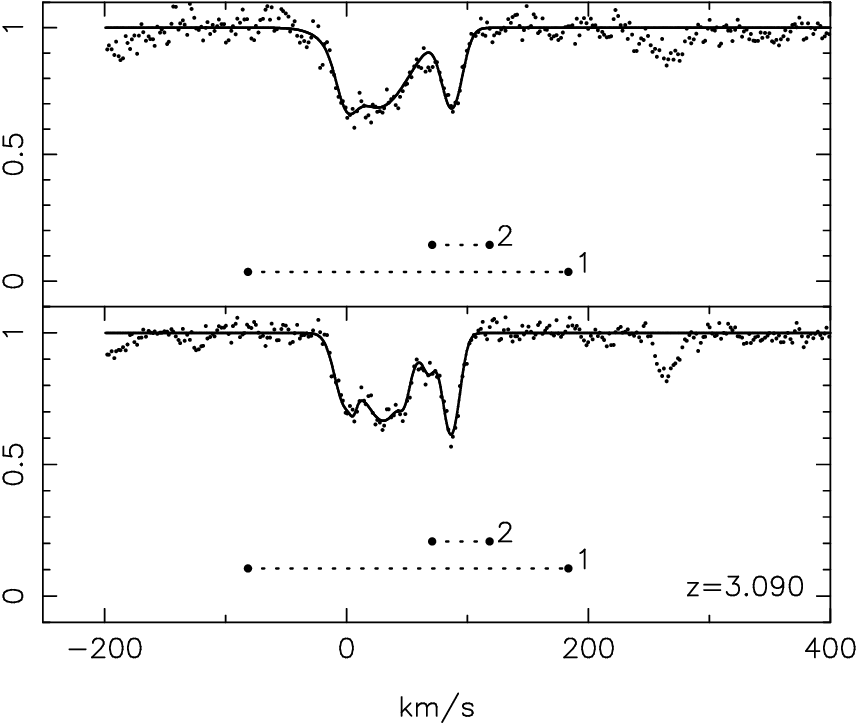}
  \includegraphics[width=5.5cm,angle=-90]{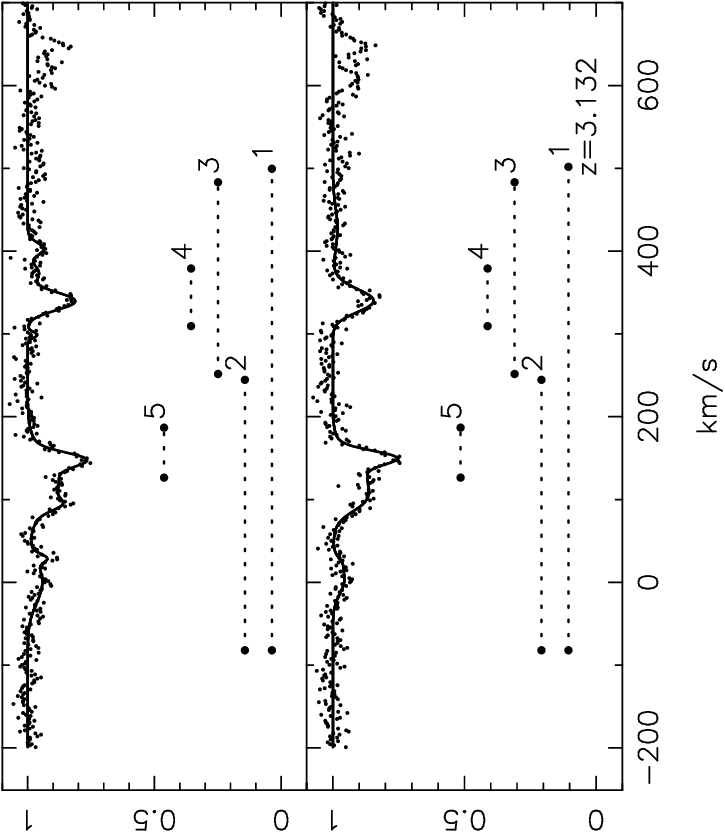}
  \includegraphics[width=5.5cm, angle=-90]{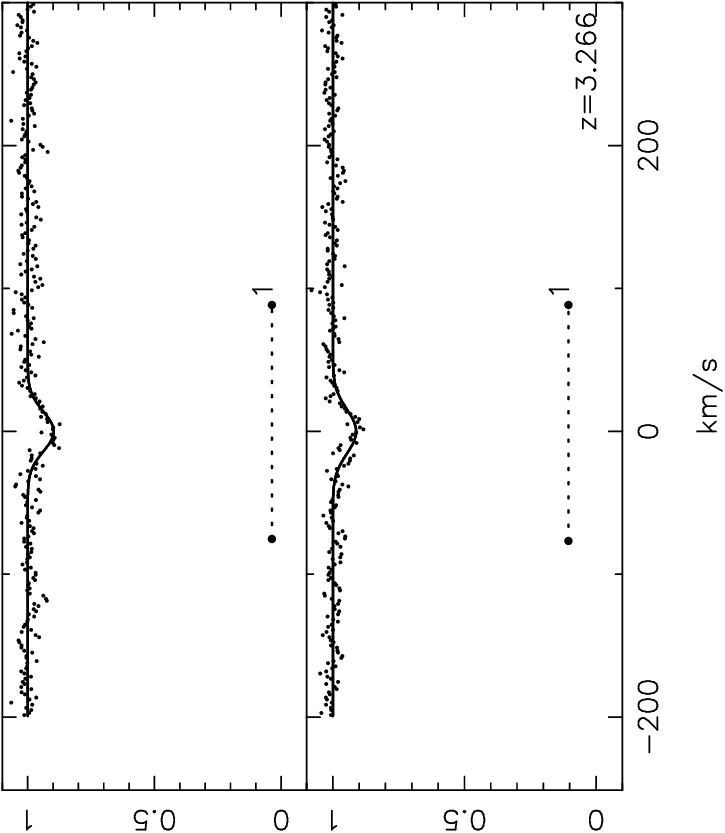}
  \includegraphics[width=5.5cm,angle=-90]{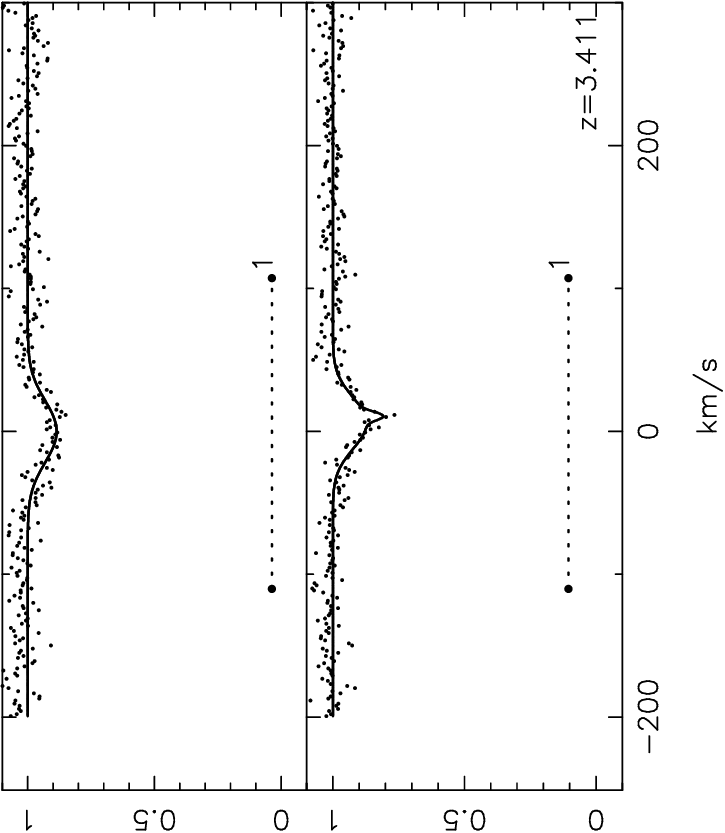}
  \includegraphics[width=5.5cm, angle=-90]{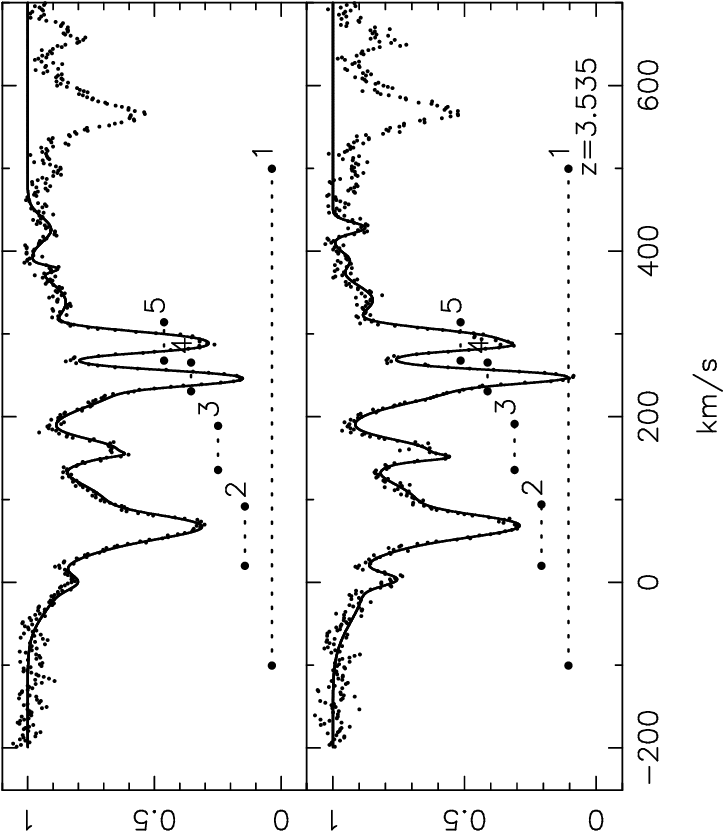}
  \includegraphics[width=5.5cm,angle=-90]{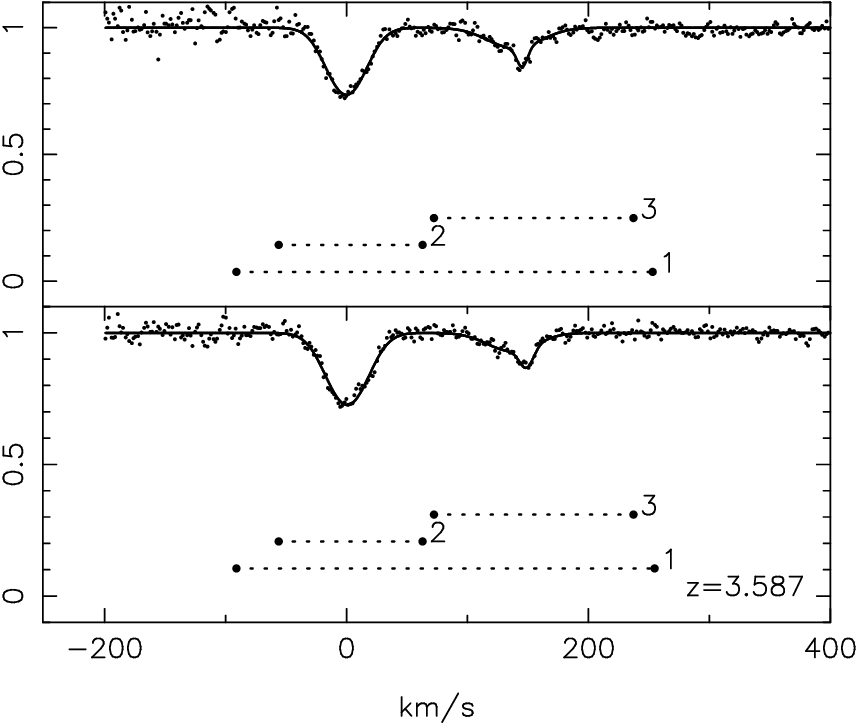}
     \caption{\small CIV systems redward of the \op emission line in the A and C lines of sight
towards Q1422+231 (continued)
\label{CIV_1422b}}
\end{figure}

\begin{figure}
     \centering
  \includegraphics[width=5.5cm, angle=-90]{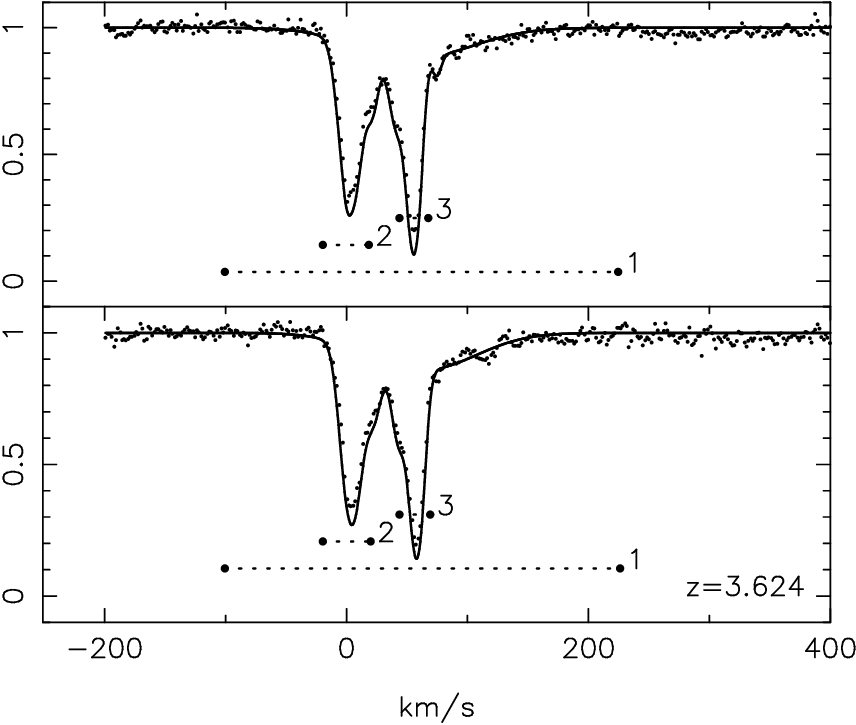}
     \caption{\small CIV systems redward of the \op emission line in the A and C lines of sight
towards Q1422+231 (continued)
\label{CIV_1422c}}
\end{figure}

\begin{figure}
     \centering
  \includegraphics[width=5.5cm, angle=-90]{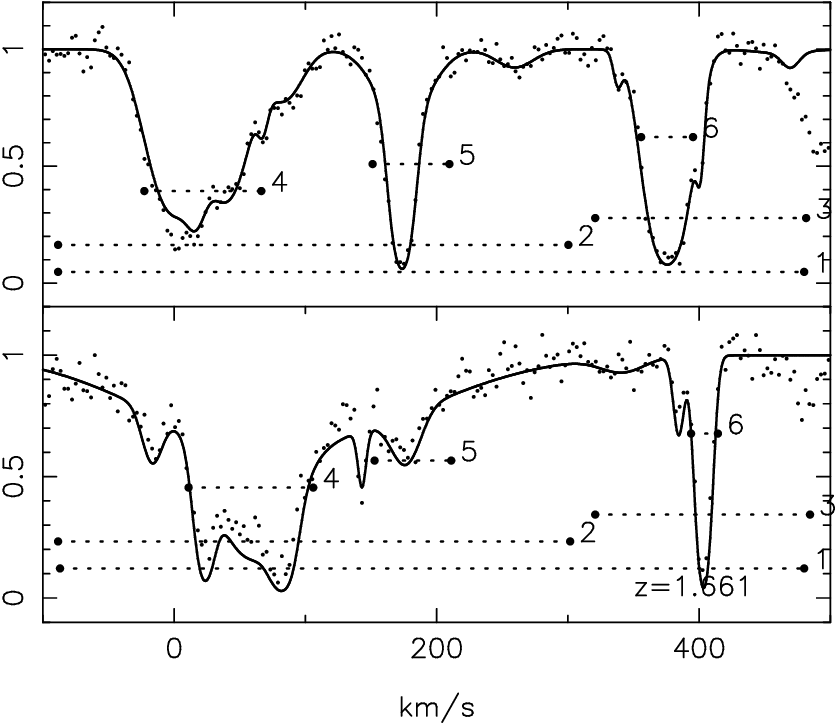}
  \includegraphics[width=5.5cm,angle=-90]{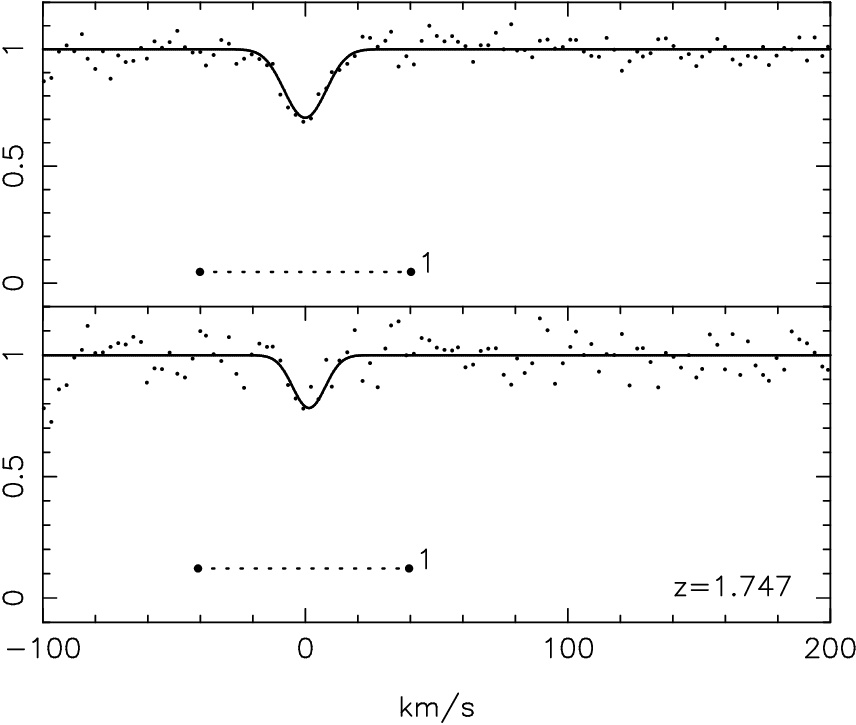}
  \includegraphics[width=5.5cm, angle=-90]{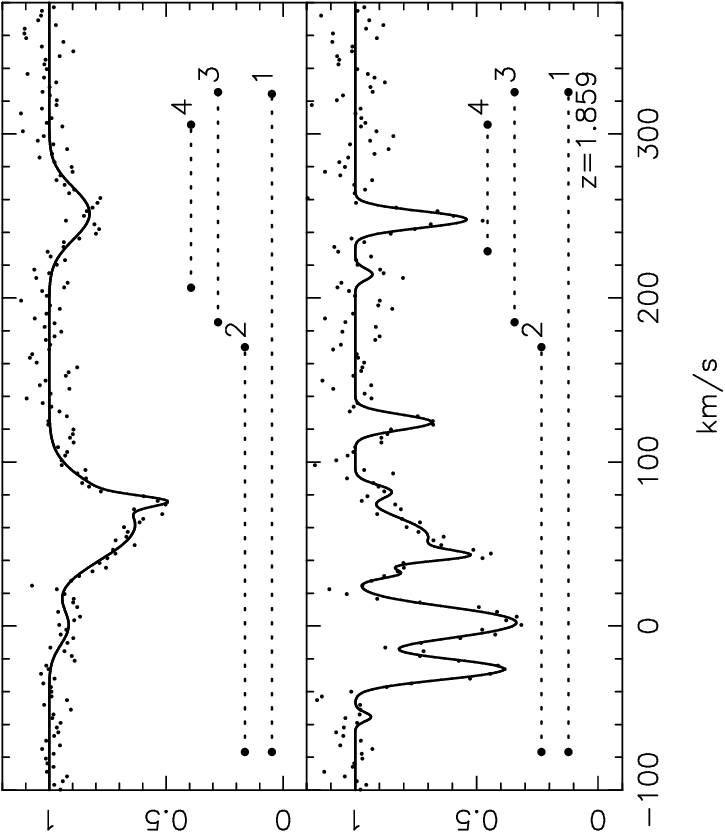}
  \includegraphics[width=5.5cm,angle=-90]{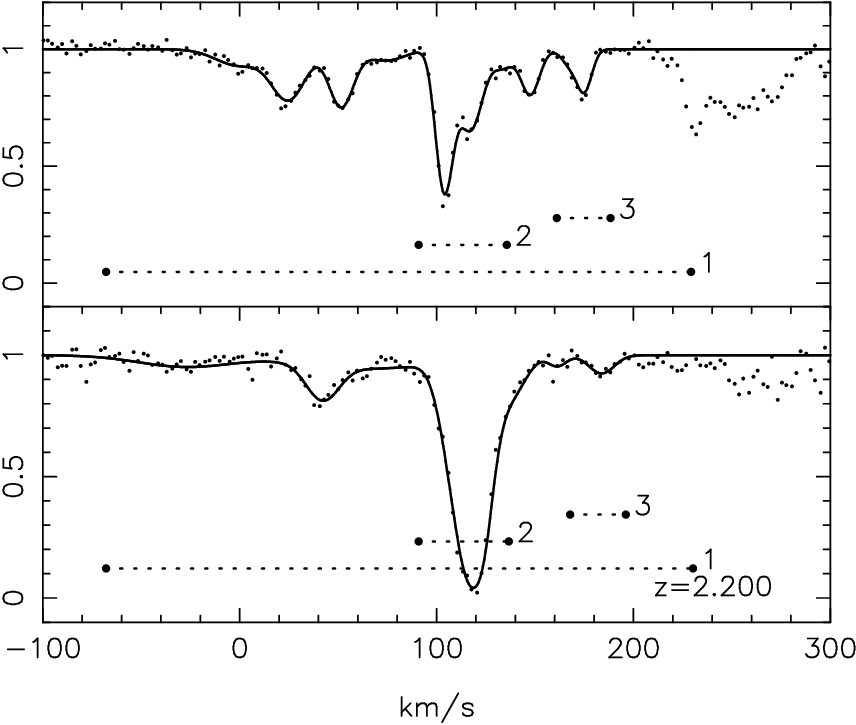}
  \includegraphics[width=5.5cm, angle=-90]{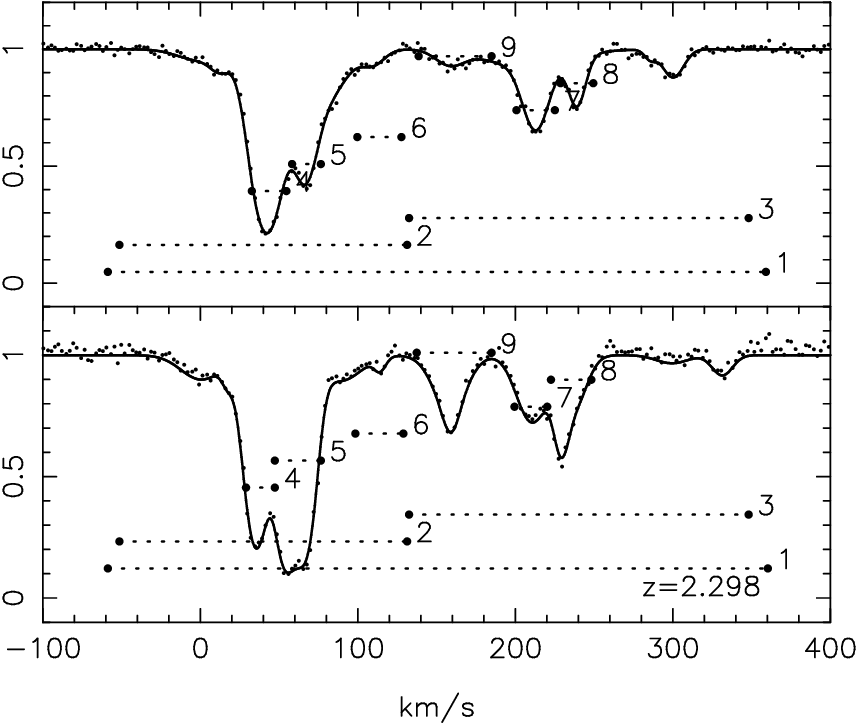}
  \includegraphics[width=5.5cm,angle=-90]{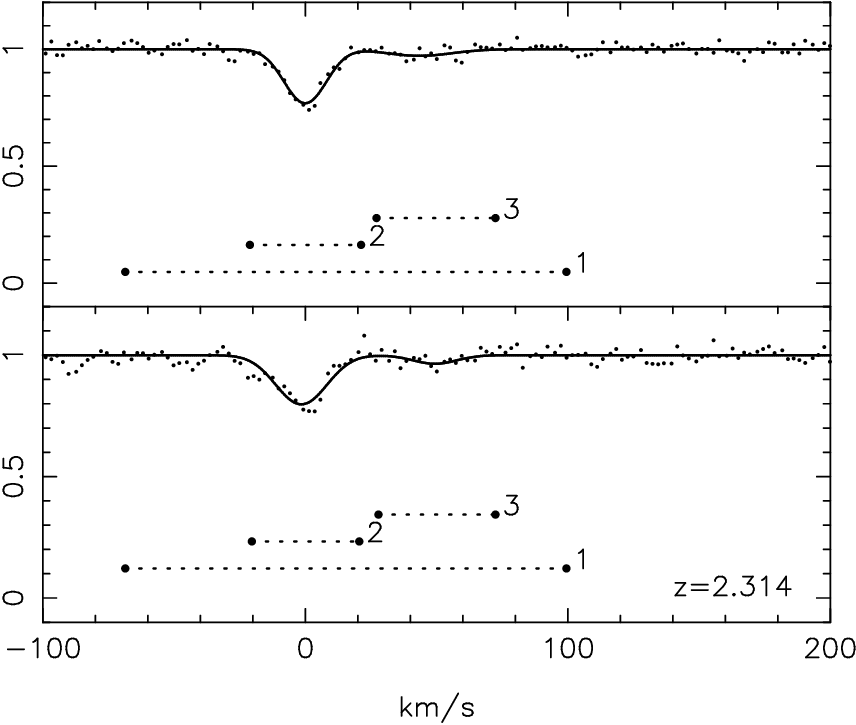}
     \caption{\small CIV systems redward of the \op emission line in the A and B lines of sight
towards HE1104-1805
\label{CIV_1104}}
\end{figure}

\begin{figure}
     \centering
  \includegraphics[width=5.5cm, angle=-90]{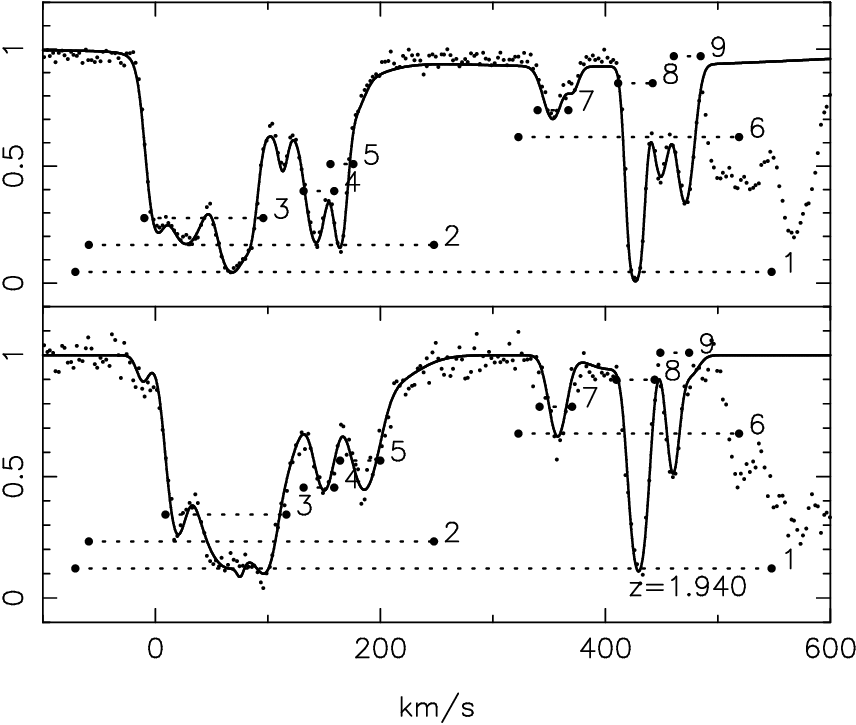}
  \includegraphics[width=5.5cm,angle=-90]{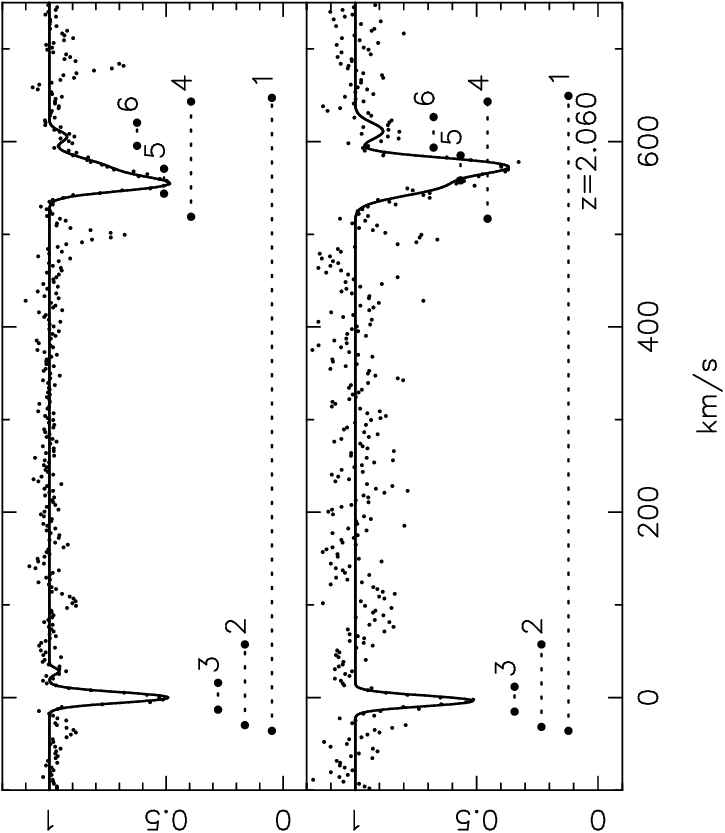}
  \includegraphics[width=5.5cm, angle=-90]{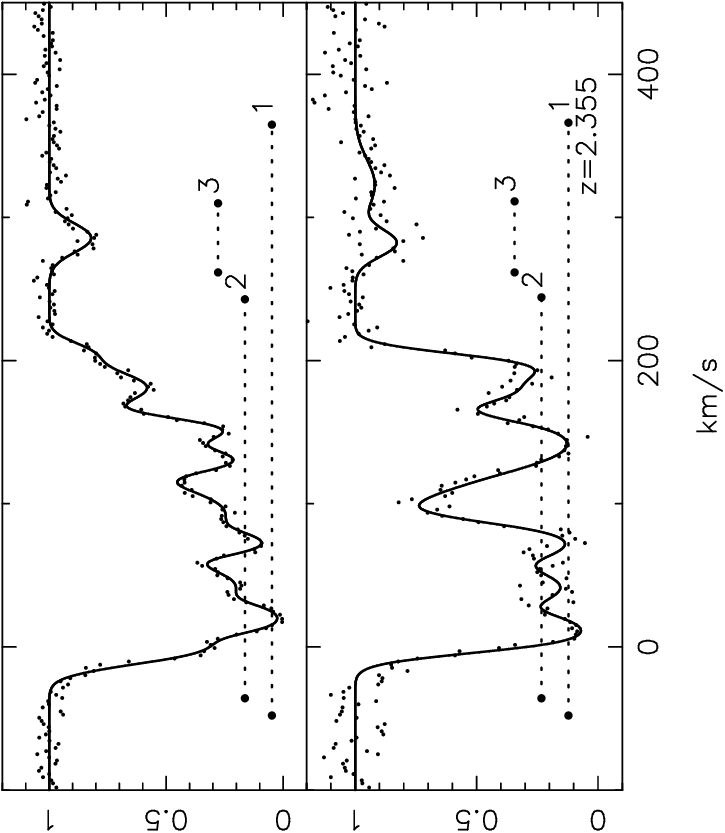}
  \includegraphics[width=5.5cm,angle=-90]{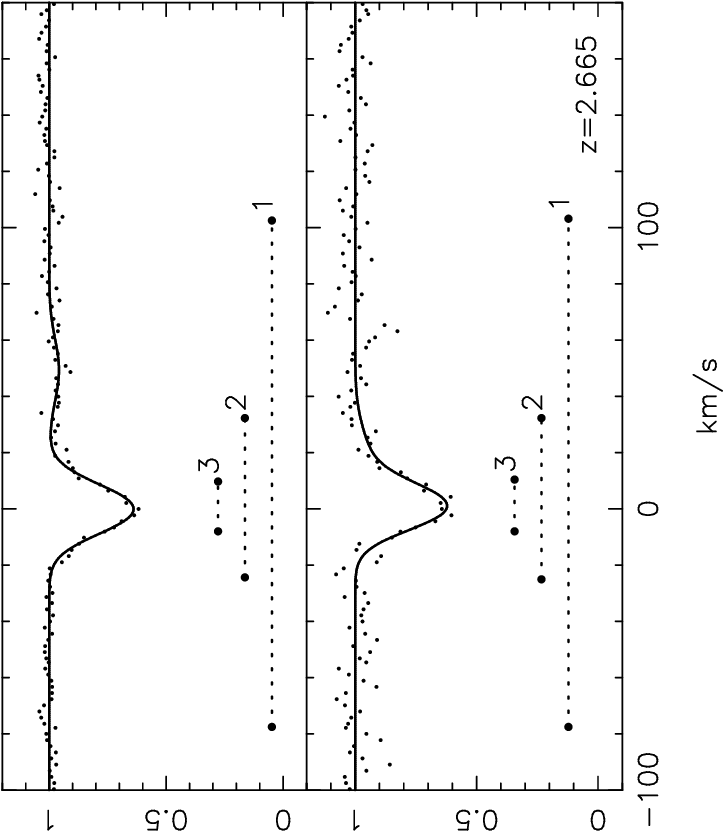}
  \includegraphics[width=5.5cm, angle=-90]{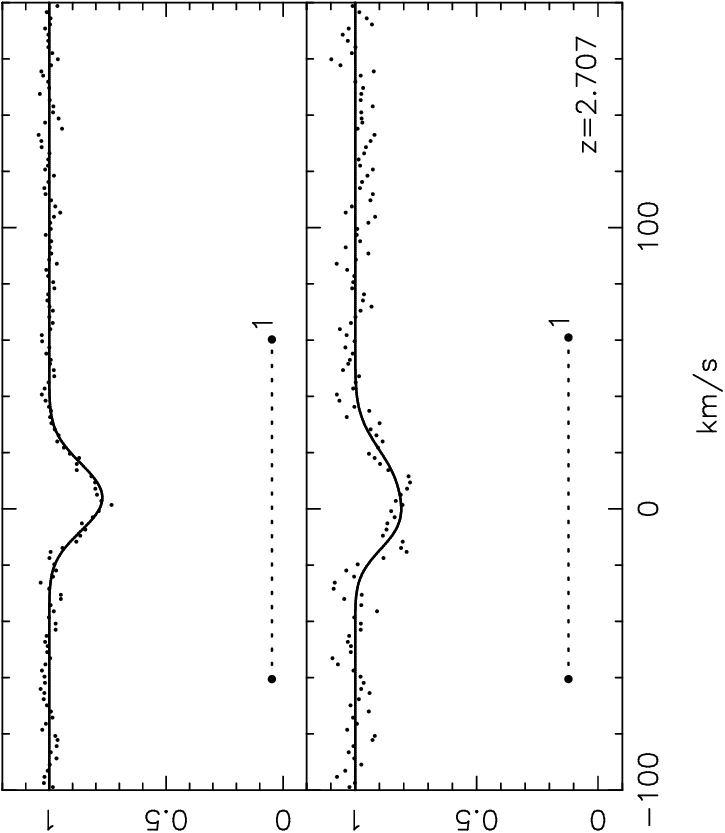}
  \includegraphics[width=5.5cm,angle=-90]{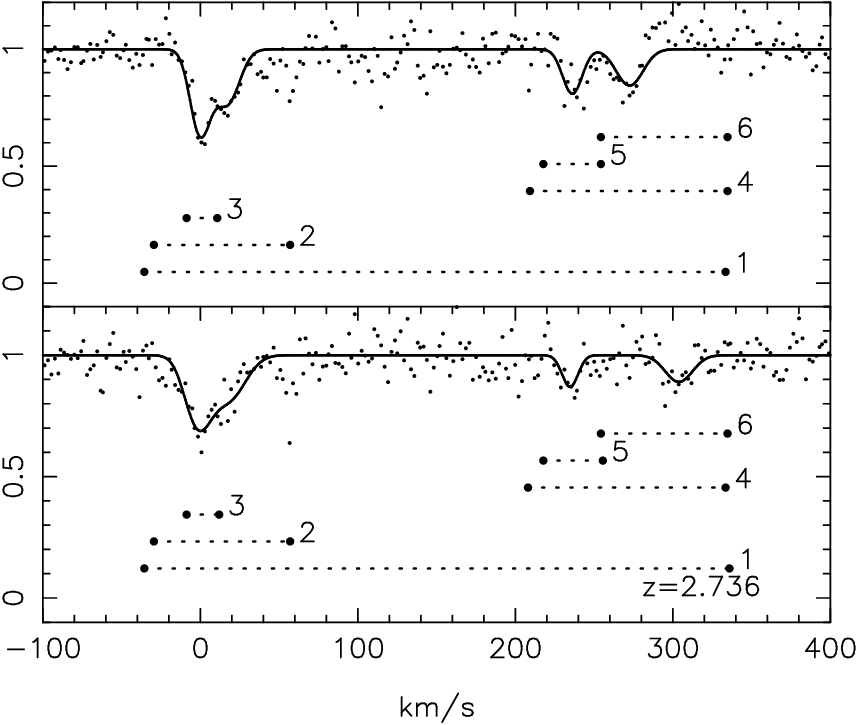}
     \caption{\small CIV systems redward of the \op emission line in the A and B lines of sight
towards UM673
\label{CIV_UM673}}
\end{figure}

The CIV lines were modelled with Voigt profiles, using the $\chi^2$
minimization fitting program VPFIT (Carswell et al 1992).  The two
lines of the CIV ($\lambda\lambda$ 1548,1550 \AA ) doublet were
profile-fitted simultaneously with as few components as necessary to
achieve a statistically acceptable reduced $\chi^2$ value (e.g.,
Carswell et al 1991).  Absorption line interlopers from other, unrelated metal
transitions were also fitted.  Then a velocity window was run over the
CIV line list to identify absorption {\em complexes}. Two absorption
components were defined to belong to the same complex, when they were
less than 1000 kms$^{-1}$ apart in their rest frame.  Then, stretches
of the spectrum with length 1000 kms$^{-1}$, each containing one CIV
system or complex, were reconstructed from the line lists.  These
'noise-free' mini-spectra where used for further analysis. They are
shown in figs.  \ref{CIV_1422}-\ref{CIV_1422c} for Q1422+231, and
figs.  \ref{CIV_1104} and \ref{CIV_UM673}, for HE1104-1805 and UM673,
respectively.

We next describe  the
velocity and column density structure of these systems, and the
dependence of these physical parameters on the line of sight separation.
Then, we attempt to measure the typical size of the absorbers.

\subsection{Complexes, subclumps, and single components}

As indicated in the above discussion there may  be reason for
subdividing the observed absorption pattern into smaller units.  To
make further progress we have attempted to crudely classify absorption
features and substructure within a given absorption complex by eye.
Once we have established a correspondence between features seen in
multiple lines of sight, we can measure the gradients in velocity and column
density across the lines of sight.

We proceed such that groups and substructures of absorption systems
thought to correspond to each other are selected interactively in both
lines of sight. The properties (total column density; column density-weighted 
velocity;
total velocity width) are then computed for each feature in both lines of sight.
We fairly arbitrarily define three classes of absorption pattern, which we
compare separately:

\medskip

{(1)  the whole absorption system-- ("Complexes")}

{(2) distinctly different groups of components within a system-- ("Subclumps")}

{(3) single absorption lines -- ("Components")}
\medskip

The meaning of class (1) and (3) is immediately clear, but class (2) is
more vaguely defined as a close subgroup within an absorption complex
-- implying that there are additional components at larger separation
than the typical intercomponent distance of the subgroup. In other
words, the transmitted flux level {\em between} clumps is substantially
higher than within them.  This classification is illustrated in figs.
\ref{CIV_1422} -- \ref{CIV_UM673}, and in tables 1-3 . The figures show
all CIV systems detected in the three QSO spectra, with the spectrum of the
A image
on top and that of the  B (or C) image at the bottom.  Redshifts are given
in the RHS bottom corner.  The velocity range of each set of components
as selected by eye using the criteria given above is indicated by a
dotted line, bounded by two larger dots.  Regions with the same numbers
in different images are meant to correspond to each other. Most of
the time the correspondence in the two spectra between the larger
complexes is unambiguous and the velocity position and width of the
window is very similar in both images. Occasionally, especially in the
case of single components, it is unclear whether we are really seeing
the same structure in the two spectra; this must be kept in mind for
later interpretation.  The numbers within a given absorption system
correspond to the entries in tables 1-3. The tables give for each
system first the redshift $z_{\mathrm abs}$  and transverse separation $r$
between the lines of sight, followed by as many entries as there are
velocity regions defined.  The parameters in each line are:  the
integrated column densities $N_{A}$, $N_{B}$ (for the A and B
images), the logarithmic column density difference $\Delta \log N$
between the two lines of sight, the difference between the column
density weighted mean velocities, $\overline{v_A} - \overline{v_B}$,
and the total bulk motion velocity widths $\Delta v_A$, $\Delta v_B$ of
each structure. If the feature can be fitted by a single Voigt profile,
no velocity width is given. The last column in tables 1 - 3 consists of
three subcolumns, giving the class of each subgroup in the window. For
example $1\ 0\ 0$ means that the window includes the whole absorption
system (class 1), whereas $1\ 0\ 3$ would imply that the system belongs
to class 1 and 3 simultaneously (i.e., it the whole system consists of
a single component). The combination $0\ 2\ 0$ denotes a subgroup of
several components.

\medskip

\begin{figure}
     \centering
  \includegraphics[width=8.5cm, angle=-90]{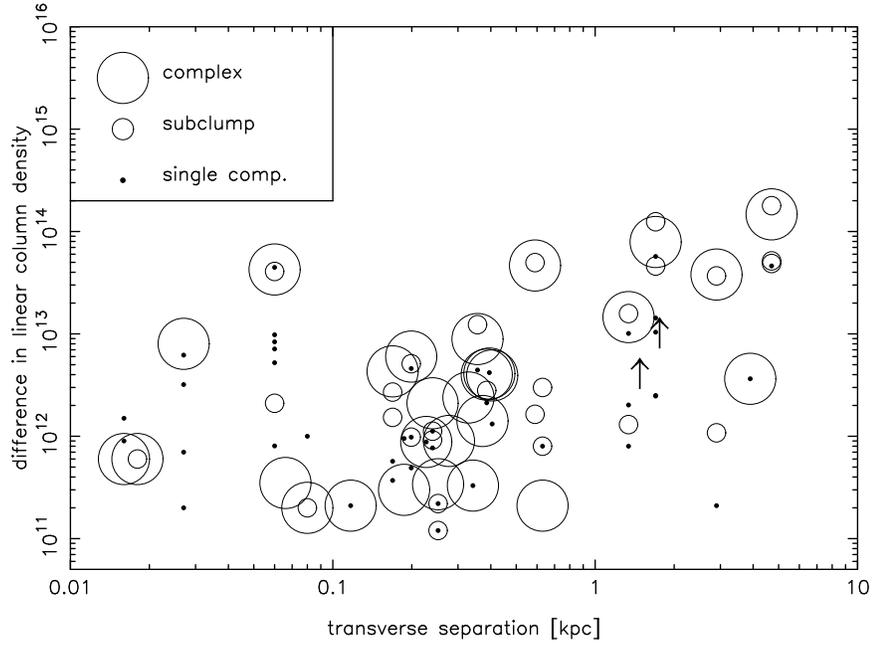}
     \caption{\small  Absolute change in CIV column density as a function
of transverse separation between the lines of sight, assuming $z_{\mathrm lens}({\mathrm 
HE}1104-1805$)=0.73.
\label{civvnstat2}}
 \end{figure}

\begin{figure}
     \centering
  \includegraphics[width=8.5cm, angle=-90]{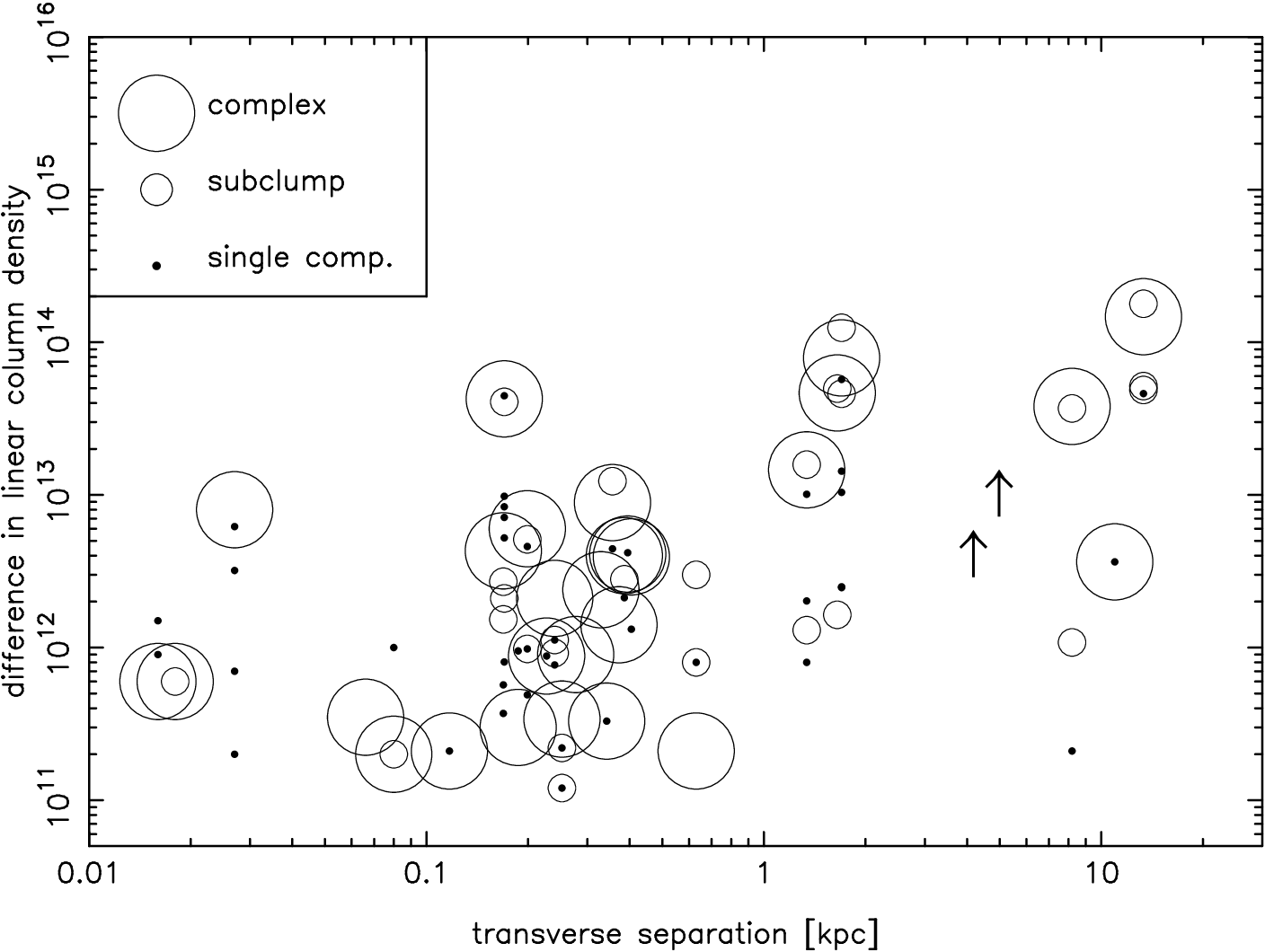}
     \caption{\small  As fig. \ref{civvnstat2}, but now assuming $z_{\mathrm lens}({\mathrm 
HE}1104-1805)$=1.32.
\label{civvnstat2aa}}
 \end{figure}

\begin{figure}
     \centering
  \includegraphics[width=8.5cm, angle=-90]{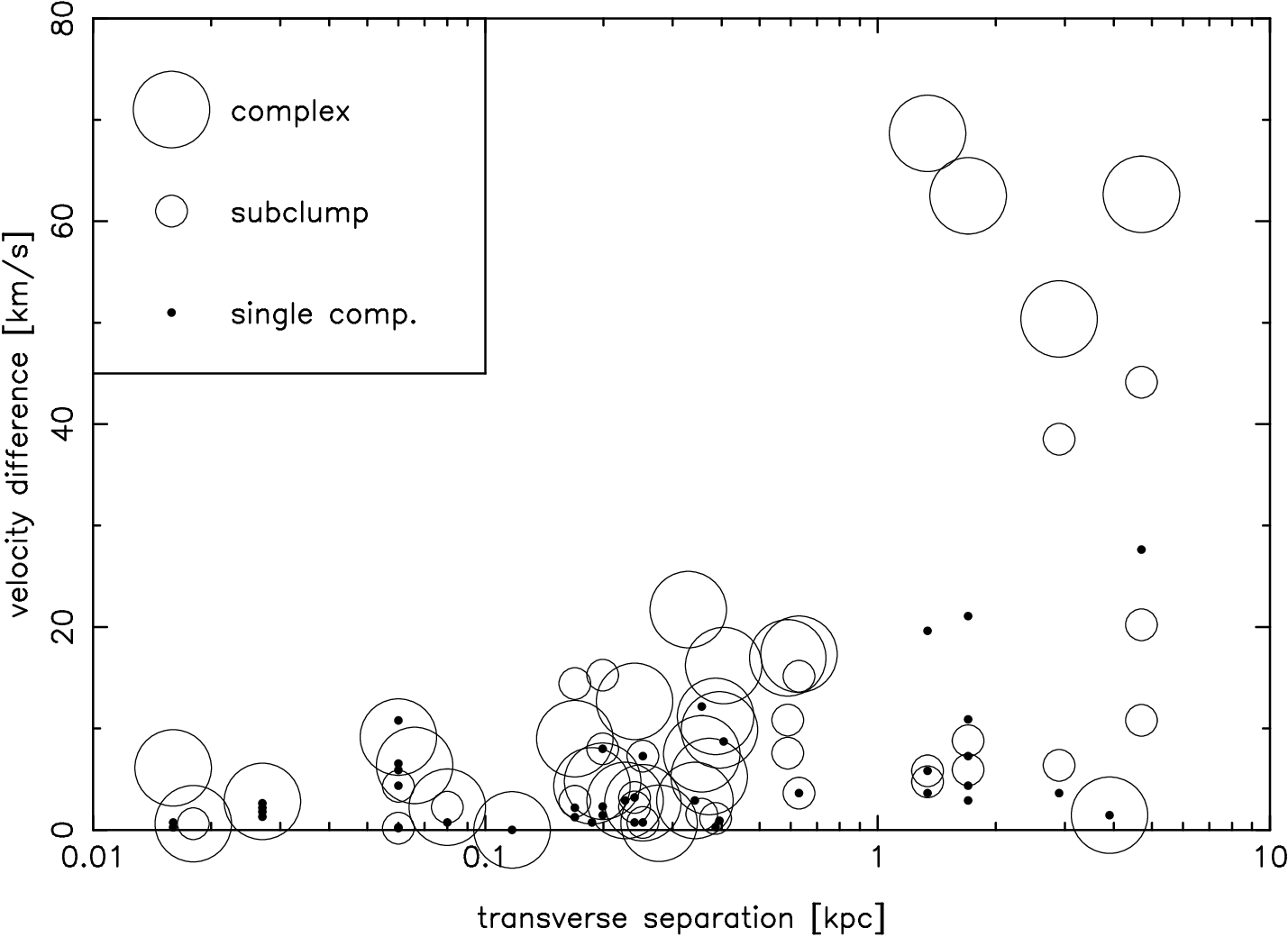}
     \caption{\small  Differences between the column density weighted velocities
along the line of sight, as a function of beam separation (assuming 
$z_{\mathrm lens}({\mathrm HE}1104-1805)$ = 0.73).
\label{civvnstat1}}
\end{figure}

\begin{figure}
     \centering
  \includegraphics[width=8.5cm, angle=-90]{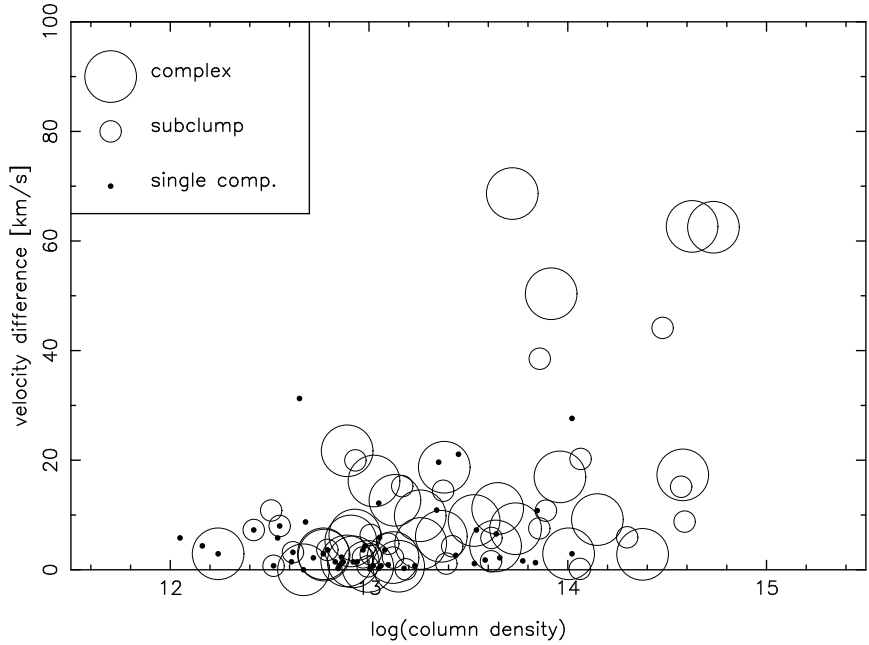}
     \caption{\small  Differences between the column density weighted velocities
along the line of sight, as a function of the column density (assuming 
$z_{\mathrm lens}({\mathrm HE}1104-1805)$ = 0.73).
\label{civvnstat3}}
\end{figure}

The results of these measurements are illustrated in Figs. \ref{civvnstat1a} --
\ref{civvnstat3}. Because the sample is very small we refrain from
imposing a uniform selection according to the signal-to-noise level or
minimum column density, and show all the data in one plot for all the 
systems detected.

\subsection{Differences in column density as a function of
separation between the lines of sight}

We can define the 'size' or 'coherence length' of a gas cloud
as the distance over which there are significant changes in the physical
parameters. The quantities most straightforward to measure are changes
in the column density and in the projected velocity between the lines of sight. 

Fig. \ref{civvnstat1a} shows the fractional difference between the CIV column
densities,
$(N_A - N_B)/{\mathrm max}(N_A,N_B)$, versus the transverse
separations between the lines of sight $A$ and $B$, on a logarithmic
scale.  The fractional difference of any absorption system obviously
lies between 0 (system identical in both lines of sight) and 1 (system
only present in one line of sight, not in the other).  The information
is given separately for all three categories as defined above, namely
absorption complexes (large circles), subclumps (small circles), and
single components (dots). The two arrows show upper limits for two
complexes where the absorption system was present only in one line of
sight but not in the other.  The beam separation is computed assuming
$q_0=0.5$ and the following lens redshifts:  UM673: $z_{\mathrm lens}$=0.493
(Surdej et al 1988); Q1422+231:  $z_{\mathrm lens}$=0.338 (Kundic et al 1997);
HE1104-1805:  $z_{\mathrm lens}$= 0.73 (Wisotzki et al.  1998). On the basis of
a time-delay estimate the latter authors have argued that the redshift
of the lensing galaxy is not as high as originally assumed
($z_{\mathrm lens}$=1.32 or 1.66, the location of two candidate absorption
systems in the line of sight) but may coincide with a system at
$z_{\mathrm lens}=0.73$. (After the present paper was completed Lidman et al
(2001) have detected a z=0.73 galaxy in emission and presented
additional arguments that this is the lensing object). The beam
separation  for an absorption system at redshift z = 2 increases by a
dramatic factor 5 when the  lens redshift is moved from 0.73 to 1.66,
so knowing the precise lens redshift is crucial for quantitative
inferences. For comparison, we show in fig.\ref{civvnstat1aa} the same
plot with the lens redshift of HE1104-1805 set to z =1.32.  The lens
redshift of UM673 is also somewhat uncertain (there are three MgII absorption
systems near $z\sim 0.5$, one of which ($z_{\mathrm abs}=0.56358$) is stronger and more symmetric
in both lines of sight than the absorption pattern caused by the
presumed lens.

\subsubsection{Testing the lensing hypothesis}

Although the scatter in figs. \ref{civvnstat1a} and \ref{civvnstat1aa}
is quite large (which is partly due to the pixel noise in the data - a
problem for weak systems close to the detection threshold), one of the
most basic trends expected is indeed observed: the larger the 
beam separation, the greater is the column density difference between
corresponding absorption systems.  This trend shows that the three QSOs
are indeed gravitationally lensed and do not consist of pairs of
individual QSOs. In the lensed case, for absorption redshifts
$z_{\mathrm lens}<z_{\mathrm abs}<z_{\mathrm QSO}$, the beam separation is a rapidly
decreasing function of redshift, whereas for genuine QSO pairs the beam
separation increases with redshift and levels off or converges only
weakly at higher redshifts, depending on the cosmological model.  Plots
of this sort should provide a useful test for checking the lens-versus-pair 
hypothesis for multiple QSO images. If, in the future, a 
sufficient number of lensed QSOs can be observed to establish good limits on 
the CIV
absorption sizes, we may even be able to reconstruct the lens redshifts
of individual lensed QSOs by matching the degree of difference between
absorption systems along their lines of sight.

\subsubsection{A mimimum size for CIV 'clouds'}

\begin{figure}
     \centering
  \includegraphics[width=8.5cm, angle=-90]{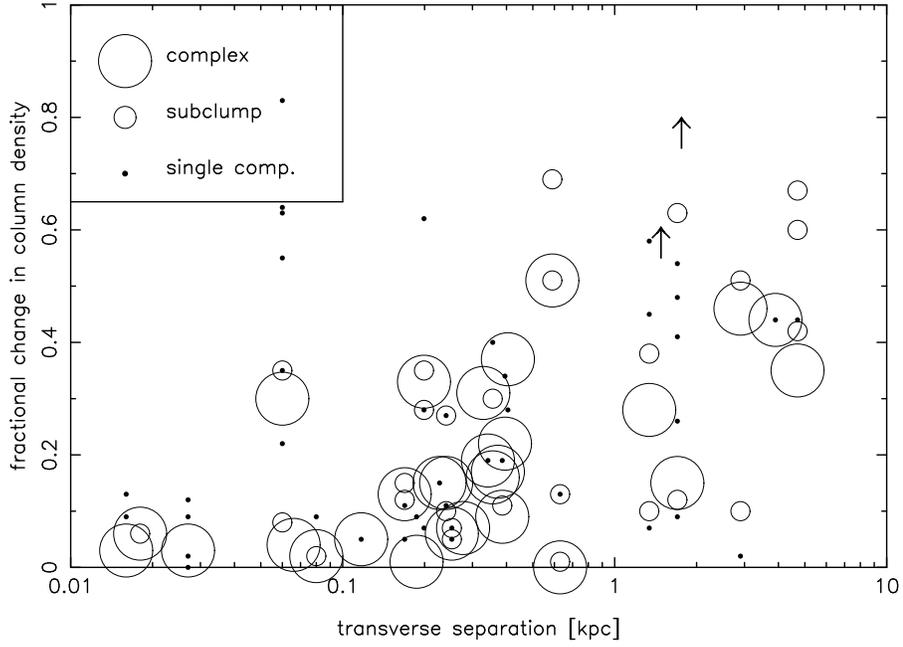}
     \caption{\small Fractional change in CIV column density as a function
of transverse separation between the lines of sight. For HE1104-1805, a lens 
redshift of 0.73 was assumed.
\label{civvnstat1a}}
\end{figure}

\begin{figure}
     \centering
  \includegraphics[width=8.5cm, angle=-90]{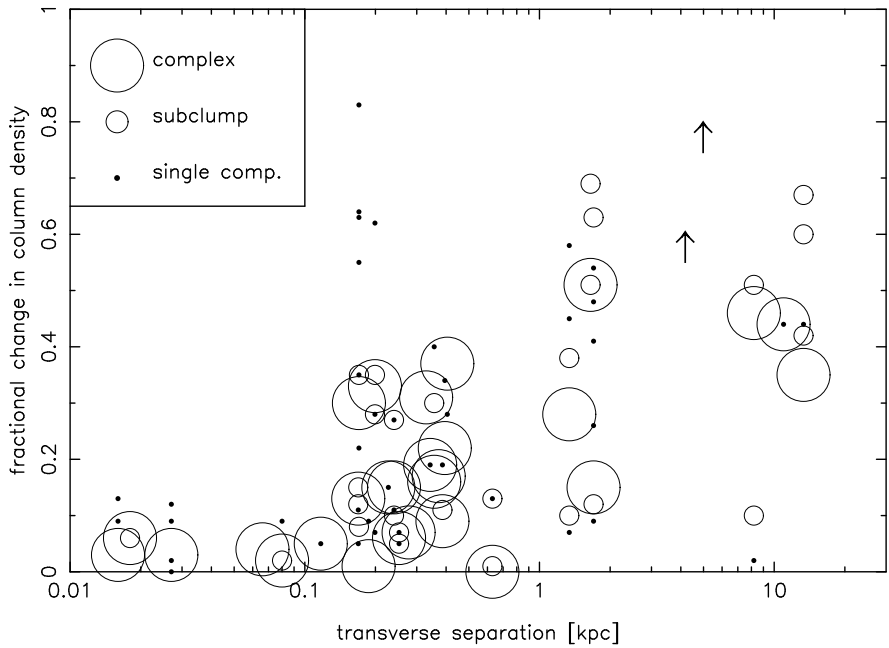}
     \caption{\small  As in fig.\ref{civvnstat1a}, but now assuming that the
lens of HE1104-1805 is at $z=1.32$.\label{civvnstat1aa}}
\end{figure}

With the exception of the absorption complex at z=2.298 in HE1104-1805
(which lies at separation 60 pc for a lens redshift z=0.73 (fig.
\ref{civvnstat1a}), but at 170 pc for a lens redshift z=1.32
(fig.\ref{civvnstat1aa}), there is very little difference between the
column densities for line of sight separations below about 300 pc
(Fig.\ref{civvnstat1a}). For separations larger than that, the
differences increase dramatically and reach 50\% just below 1 kpc.  
{\em Apparently,
the spatial distribution of highly ionized CIV gas is mostly
featureless on scales below $\sim 300 pc$}. At the same time, detectable
CIV absorption extends over at least several kpc. These
result have important consequences for the nature of CIV absorbers which
we will explore further below.  Fig.\ref{civvnstat2} and
\ref{civvnstat2aa} give the analogous plots for {\em absolute}
differences in column density.

\subsection{Relative differences between  
velocities along the line of sight as a function of
separation}

Fig. \ref{civvnstat1} shows the velocity 'shear' as a function of beam
separation. The shear or velocity difference $\Delta v$ between the
lines of sight is defined as the difference between the column density
weighted velocities, $\Delta v = \overline{v_A} - \overline{v_B}$, as seen in the two spectra.
The shear slowly rises up to a transverse separation 200-300 pc, beyond
which it increases rapidly to $\sim 60-70 $kms$^{-1}$, at several kpc.  Such
values are not unexpected for clouds which partake in the
velocity dispersion even of a dwarf galaxy. However, the values should
be taken as upper limits, since we know from the results in the previous
paragraph that the column densities also have a  coherence length
on a similar scale. Because of the column density weighting, we may
partly confuse column density differences with velocity differences.

\subsection{The relative differences between  
velocities along the line of sight as a function of the column density}

In fig.\ref{civvnstat3} $\Delta v$ is plotted versus the CIV column
density of the system. The observed increase of $\Delta v$ with N(CIV) probably
implies that the highest density gas experiences the largest
velocity dispersion and that the gas is more quiescent where the column 
density is low. 

\subsection{A lower limit on the overall size of CIV systems}

A crude  estimate of the total sizes of CIV complexes, $L_{\mathrm tot}$, based
on the sizes of individual CIV components
can be made as follows:

Multiplying the scale $r_{1/2}$ over which the individual absorption components
change by 50\% in column density, say, $r_{1/2}\sim 300$ pc, by a factor 4 we
get an estimate of the minimum size (2 FWHM) of a cloud giving
rise to a single CIV component. Then the full CIV complex must be at
least as extended along the line of sight as indicated by the number of 
components $N_c$ times the
size of each component, i.e., 
\begin{eqnarray} 
L({\mathrm tot)}\sim 1.2  N_c f^{-1} \left(\frac{r_{1/2}}{300\mathrm pc}\right) 
{\mathrm kpc},
\end{eqnarray} 
where $f$ is the filling factor of the CIV system for individual
clouds.  For example, a CIV system with 5 individual components with
basically no intercloud space ($f\approx 1$) would imply a size of 6 kpc along the
line of sight. If CIV clouds form in situ (i.e., the material ending up
in the clouds at density $\rho_c$ was derived from their immediate
surroundings (at original density $\overline{\rho}$) then the formation
of the cloud produces an intercloud space such that $f^{-1} =
\rho_c/\overline{\rho}$. Clouds with an overdensity
$\rho_c/\overline{\rho}= 10$ with respect to the background medium
would then have a filling factor $f=1/10$ and, according to the above
equation, the CIV system would have a typical size of $60$ kpc.

\subsection{Maximum likelihood size estimates for CIV complexes}

We use a maximum likelihood method to obtain a parametric estimate
of the column density gradients in the plane of the sky
given the column densities for a number of systems in both lines of sight or 
the column density in one and an upper limit in the other line of sight. 
This is a
generalization of the standard technique for estimating the sizes of
absorption systems from the numbers of ``hits'' and ``misses'' (McGill
1990). A similar approach has been discussed by Smette et al. (1995) to
derive lower limits to \op cloud diameters. Here we can directly
operate with actual column density measurements, or with the upper limits in
cases where a system was detected only in one of the two lines of sight.
Each absorption system yields only one pair of column
density measurements, but if we make the (doubtful) assumption that all
clouds are identical, taking one measurement for many systems is
equivalent to making many measurements of a single system.

\begin{figure}
     \centering
  \includegraphics[width=11.5cm, angle=-90]{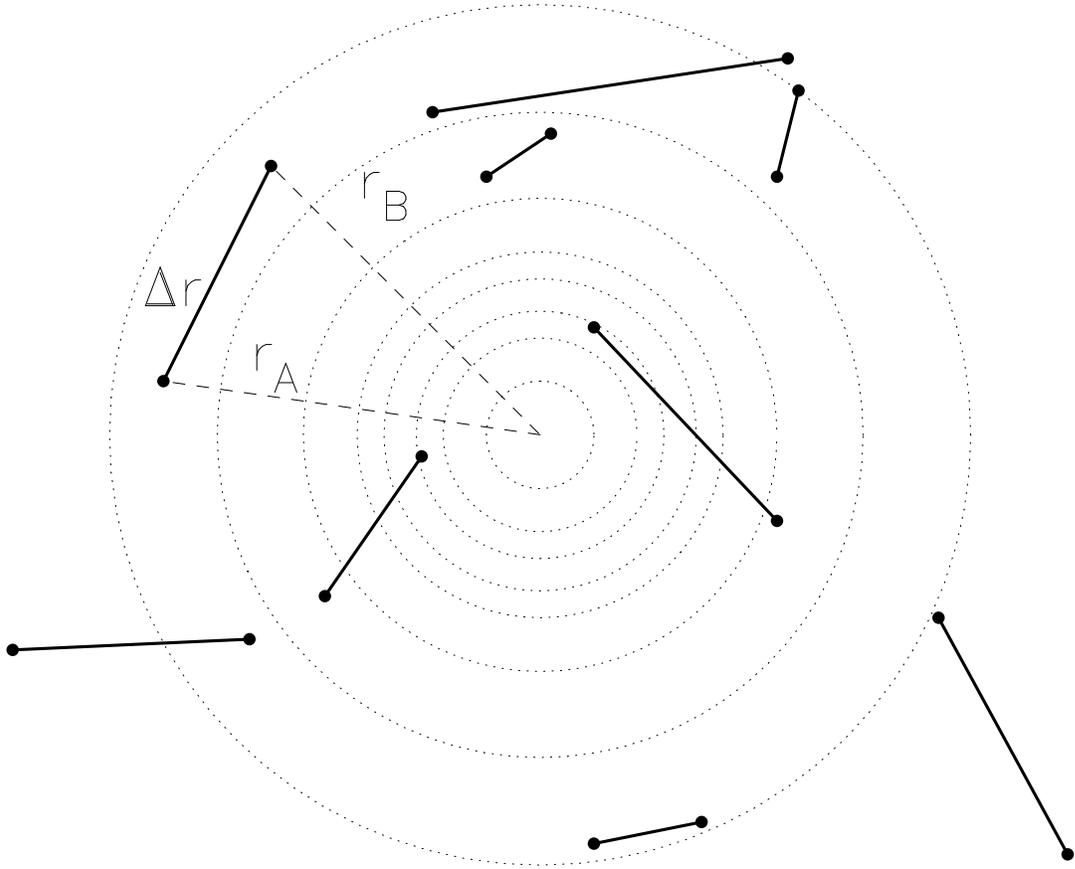}
     \caption{\small diagram illustrating the statistical reconstruction of the 
column density
contours (dashed lines) of a simple cloud model using pairs of column density 
measurements from random hits by double lines of sight.\label{contours}}
 \end{figure}

For those systems with
definite column densities in both lines of sight we compute 
the conditional
probability density  $p(N_B | N_A, \Delta r)$, of measuring a column
density $N_B$ in a given system in line of sight B, given a column 
density $N_A$ in line of sight $A$ and given a transverse separation 
$\Delta r$ between
the lines of sight.  The probability follows from simple geometric arguments if
one assumes that all absorbers possess spherical symmetry and have
identical column density profiles, with a semi-definite density
gradient, $dN/dr < 0$ (the assumed situation is illustrated in
fig.\ref{contours}).   Then, the impact parameter $r_A$ of line of sight $A$
from 
the center of an absorbing clump is related to the impact parameter 
$r_B$ of line of sight $B$, the line of sight transverse
separation $\Delta r$, and the angle $\phi$ between $r_A$ and $\Delta r$
simply by the cosine law
\begin{eqnarray}
r_B^2 = r_A^2 + (\Delta r)^2 - 2 r_A\Delta r \cos \phi.
\end{eqnarray}
A uniform random distribution in $\phi$ (given fixed $r_A$ and $\Delta r$)
then generates the conditional probability in $r_B$ and thus in $N_B$
\begin{eqnarray}
p(N_B | N_A, \Delta r) dN_B = \frac{-r(N_B)}{\pi r(N_A)\Delta 
r}\frac{dN_B}{\sqrt{1 - \left(\frac{(\Delta r)^2 + r(N_A)^2
- r(N_B)^2}{2r(N_A) \Delta r }\right)^2}}\frac{dr}{dN}(N_B),
\end{eqnarray}
where the functions $r(N)$ and $(dr/dN)(N)$ are derived from a simple model.
Here we adopt a Gaussian column density profile,
\begin{eqnarray}
N = N_0 \exp\left(- \frac{1}{2}\left(\frac{r}{r_0}\right)^2\right),
\end{eqnarray}
where the central column density $N_0$ and the Gaussian width 
$r_0$
are the desired parameters of the model. This parameterization was chosen
merely because it is well behaved at the origin, and because its inverse 
and first
derivative are easy to calculate analytically. The few data points and lack
of our knowledge about the detailed cloud physics does not appear to justify
more sophisticated modelling. It is important, though, to keep in mind that
the actual column density distribution may be neither  spherically symmetric,
or even similar from absorber to absorber. Nor may the column density
be close to the values computed from the model in column density ranges where 
there were few data points to start with, and which consequently are not
well constrained. Most importantly, the assumption of a monotonic radius-column
density relation may be wrong, and we may be measuring a typical spatial
fluctuation scale rather than the width of a coherent object.
With these cautionary remarks we proceed to derive the parameters $N_0$ and
$r_0$ from a maximum likelihood method, minimizing
\begin{eqnarray}
\ln L = \sum_{j=1}^J \ln p_j + \sum_{j=J+1}^{J_{max}}\ln\int_0^{N_{<}} p_{j} 
dN_B.
\end{eqnarray}
Here  
\begin{eqnarray}
p_j = p(N_B(j) | N_A(j),\Delta r (j), N_0, r_0)
\end{eqnarray}
is the conditional probability, now written as a function of the model
parameters $N_0$ and $r_0$, and the sum over the integrals from $J+1$
to $J_{max}$ accounts for those cases where one measurement gives only
an upper limit $N_<$ on the column density $N_B$ (Avni et al 1980). The program
was
implemented using the minimization routine AMOEBA (Press et. al 1986),
and it was tested with random pairs of data points drawn from the said
Gaussian model. It was run on those 31 absorption complexes for which
at least one line of sight had a CIV absorption system stronger than $3\times
10^{12}$cm$^{-2}$. Unfortunately, it was not always possible to get universal
convergence (= results independent of the initial values).  Likely reasons are
the smallness of the sample, and the failure of the simple model.  By
varying the initial conditions in most cases convergence to values near
$r_0$=3 kpc, $N_0$=5$\times 10^{14}$cm$^{-2}$ was obtained, but
sometimes $r_0$ was off by a factor of several. Running the same
test on the 27 groups of `single components' (as defined above)
the best values were $r_0$=19 kpc, $N_0=10^{15}$cm$^{-2}$, which gives
an idea of the large uncertainties involved. Thus, the technique clearly
confirms what we are already have seen by eye - structures in the absorbing 
gas with a coherence length
at least on the order of kiloparsecs - but currently the samples are
not adequate to justify more sophisticated analyses. 
Moreover, lines of sight with larger separation are needed since it appears 
that the largest cloud sizes have not been sampled by the present dataset.

\section{Energy input and turbulence}

We have derived size estimates for the high ionization gas clouds and
measured the degree of decoherence between the projected line of sight
velocities across the sky. Combining these results with other findings
about the CIV absorbers, we are now in a position to impose constraints
on the physical state and origin of the gas, and to compare the
findings with the properties of other, previously known galactic or
extragalactic astrophysical environments.

The degree to which gas on a given spatial scale has been stirred by
recent stellar processes should give us an idea of any recent star
formation activity (winds, SN explosions, ionization fronts). From the
data described above it is possible, in a crude way, to measure the
turbulence in the CIV gas as a function of spatial scale.  Then some
basic ideas about astrophysical turbulence (e.g. Chandrasekhar 1949 ;
von Hoerner 1951; Kaplan \& Pikelner 1970) can be used to estimate the
approximate rate of mechanical energy input $\epsilon_0$ into the gas.

\subsection{Turbulence deduced from velocity differences between the 
lines of sight}

Most often, astrophysical turbulence has been discussed in connection
with the Kolmogorov theory. Here kinetic energy is injected at a rate
$\epsilon_0$ into the gas on a certain scale determined by the geometry
and dynamics of the underlying energy source. For fully developed
turbulence, the energy propagates with a constant energy transfer rate
$\epsilon \approx \epsilon_0$ through eddies of ever smaller size,
until viscosity transforms it into heat. The usual dimensional analysis
leads to a relation between the energy transfer rate $\epsilon$ [in
units of ergs g$^{-1}$ s$^{-1}$, or cm$^2$ s$^{-3}$], and the RMS
velocity $v_s$ between points with spatial separation $s$,
\begin{eqnarray}
\epsilon\sim \frac{v_s^3}{s}.
\end{eqnarray} 
The customarily used structural
function 
\begin{eqnarray}
B(s) \approx \overline{[v(s') - v(s'')]^2}\  \simeq\ \overline{v_s^2}\ \approx 
\ (\epsilon s)^{2/3}
\end{eqnarray} 
(e.g., Kaplan \& Pikelner 1970) can 
be measured from the pairs of column density weighted, line of sight 
velocities,
as a function of projected beam separation. Here the average is taken over
all points $s'$ and $s''$ with separation $s$.

For the CIV gas the quantities $v_s$ and the projected separation can be read
from tables 1 - 3, and the energy transfer rate may be computed
and compared with measurements from other known astrophysical
environments.

At this point it is worth remembering that the {\it observed} quantity $r$ 
is the transverse separation between
two points in two lines of sight, i.e., the distance between the
points projected onto the plane of the sky; we do not
know the true separation $s = \sqrt{s_{\parallel}^2 + r^2}$.
Thus we are not really measuring $\overline{[v(s') -
v(s'')]^2}$ but $\overline{[v(r') - v(r'')]^2}$. The
relation between projected separation
$s_{\perp}$ and true separation $s$ depends on the spatial distribution
of the clouds and complicates the measurement of the structural
function (von Hoerner 1951; M\"unch 1958; Scalo 1984), leading to an 
observational
underestimate of the true value of $B(s)$ on scales where
$r<<s_{\parallel}$, i.e., when the transverse scales observed are much
smaller than the extent of the CIV clouds along the line of sight. This is 
almost
certainly the case for part of our measurements, since beam separations
less than a few hundred pc are smaller than the coherence length of
individual CIV clouds. Thus we would not expect to
observe a Kolmogorov law (a $B(r)\propto r^{2/3}$
relation) on small scales even if the spectrum of turbulence conforms to
it (i.e., $B(s)\propto s^{2/3}$).

\begin{figure}
     \centering
  \includegraphics[width=11.5cm, angle=-90]{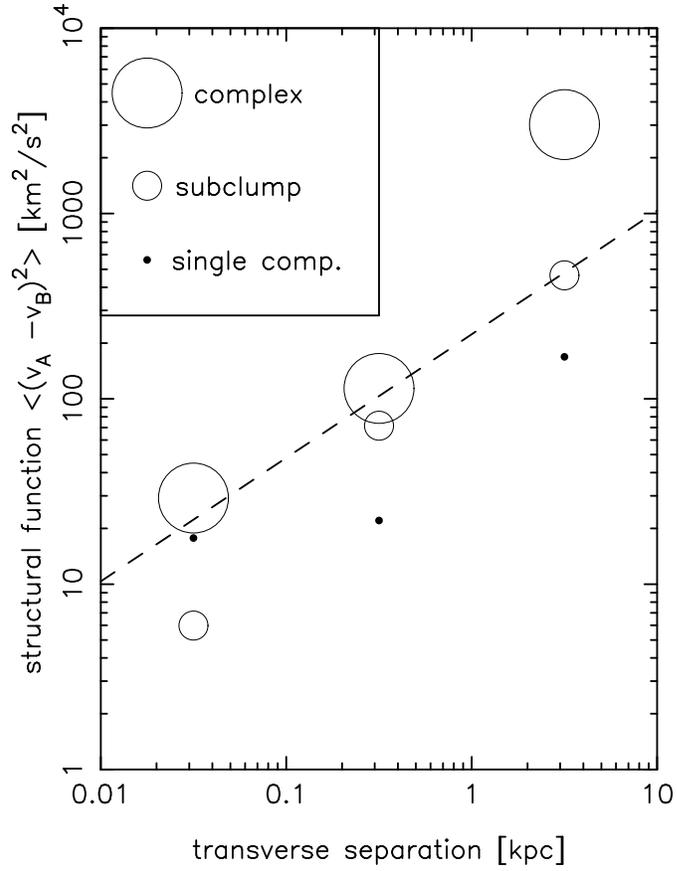}
     \caption{\small  Structure function $<(v_A - v_B)^2>$ as a function of beam 
separation. To reduce the scatter the datapoints of the previous figures have 
been collected in three bins. The dashed line illustrates a Kolmogorov spectrum 
with a vertical normalization corresponding to an energy transfer rate 
$\epsilon = 10^{-3}$ cm$^2$ s$^{-3}$.\label{strucfunc}}
\end{figure}

Keeping this complication in mind, we proceed to plot the {\it observed}
structural function $B_{\mathrm obs}(r) = \overline{[v(r') -
v(r'')]^2}$ for our sample of CIV systems as a function of the
transverse beam separation $r$ in Fig.\ref{strucfunc}.  The plot is
based on the same data points as listed in tables 1 - 3. However, now the
symbols no longer refer to individual CIV systems but give the mean
values after the data points have been collected into three logarithmic
bins ranging from 10 pc to 100 pc, 100 pc to 1 kpc, and 1 kpc to 10
kpc. For illustrative purposes the dashed line shows a Kolmogorov
spectrum ($\log B(r) = 2/3 \log r + 2/3 \log \epsilon$), normalized
such that the energy input is $\epsilon$ = 10$^{-3}$cm$^{2}$s$^{-3}$.
The expected trend for $B(r)$ to increase with $r$ is observed, but the
agreement with a single power law of slope 2/3 is not overwhelming.
There appears to be a tendency for the CIV complexes (large circle
symbols) to be more turbulent at the largest separations (or less
turbulent at the smallest separations - we do not know the correct
normalization a priori) than our naive application of the Kolmogorov
assumptions would lead us to expect. This would not be surprising if
the velocity differences between the absorption systems on the largest
scales are due to a different source of velocity shear. For instance,
in a simple hierarchical scenario for galaxy formation, the largest velocity
widths of CIV complexes are contributed by the gravitational
motions of separate protogalaxies, rather than by turbulent motion on
intragalactic scales (Rauch, Haehnelt \& Steinmetz 1997). Moreover, the
underlying assumptions of the Kolmogorov process (incompressible,
subsonic gas, no magnetic fields) may be violated by the CIV gas.
However, let us assume that there is at least some overlap between the
spatial scales covered by our sample of CIV systems and the `inertial'
range of scales where the energy transfer rate is indicative of the
original energy input. Then, judging from the departures of the data
points from a straight line (fig.\ref{strucfunc}) it appears that we
are not making an error larger than (at most) an order of magnitude in
using the y-intersect of the plot in fig. \ref{strucfunc} as a crude
estimate of the energy input rate $\epsilon_0$.  Taking values $B(r)$ =
100 km$^2$s$^{-2}$, $r = 300$ pc as reference values (the midpoint of the
power law in
fig.  \ref{strucfunc}), the energy transfer rate is 
\begin{eqnarray}
\epsilon \sim 10^{-3}{\mathrm cm}^2 s^{-3}. 
\end{eqnarray}
This is considerably less than values
measured for the Orion nebula, where $\epsilon \sim 0.1 - 1$
cm$^2$s$^{-3}$ (e.g. Kaplan \& Pikelner 1970), but it is
comparable to the global rate of energy input into the ISM of our
Galaxy.\footnote{
The value of $\epsilon$  and
the actual amount of turbulence in CIV systems could be somewhat smaller
than that derived above. 
Our ability to measure very small values of $B(r)$ ($<\, 
$a few km$^2$s$^{-2}$) is
limited by the measurement errors of $v_A$
and $v_B$. As mentioned above, there may be additional sources of velocity
shear, e.g.,  gravitationally induced bulk motions. In addition,
it is not clear how to distinguish velocity shear from density
differences across the lines of sight.  (Column) density fluctuations as caused,
e.g., by clumpiness in an otherwise laminar flow could mimic turbulent
motions.} 

\subsection{Turbulence as measured along the line of sight}

Absorption line profiles can also be used to directly probe the
turbulent velocity {\em along} the line of sight, if the line profiles of two
ions of different mass are available. Such measurements have been done
earlier for individual CIV components, using the CIV and SiIV profiles
of strong CIV systems (Rauch et al 1996). A typical value for the
turbulent line of sight RMS velocity contribution to the width of individual CIV
profiles was found to be 
\begin{eqnarray}
v_{\mathrm loS} = \sqrt{\frac{1}{2}}\, 
b_{\mathrm turb}=\sqrt{\frac{1}{2}\left(b^2 - \frac{2kT}{m}\right)}\simeq 4.5 
{\mathrm \, kms}^{-1},
\end{eqnarray}
where $b_{\mathrm turb}$ is the turbulent Doppler parameter of the CIV line, 
$b$ is the total Doppler parameter, and $T$ is the temperature.

The turbulence {\it along} the line of sight
may be compared with the turbulence {\it across} the lines of sight.  For this
comparison we need to make an assumption about the typical size of 
individual CIV
clouds. If a fiducial value of 300 pc is adopted once more,
Fig.\ref{strucfunc} gives the RMS velocity difference across the sky,
\begin{eqnarray}
\sqrt{\overline{v_r^2}}
\simeq \sqrt{B(300 \mathrm pc)} \approx 4.7 {\mathrm\, kms}^{-1} 
\end{eqnarray}
(as read from
the value of $B(r)$ for single components, in the central bin). The
good agreement between the turbulent measurements across and along the line of 
sight
indicates that the typical size adopted has the right
order of magnitude.

\begin{figure}
     \centering
  \includegraphics[width=11.5cm, angle=-90]{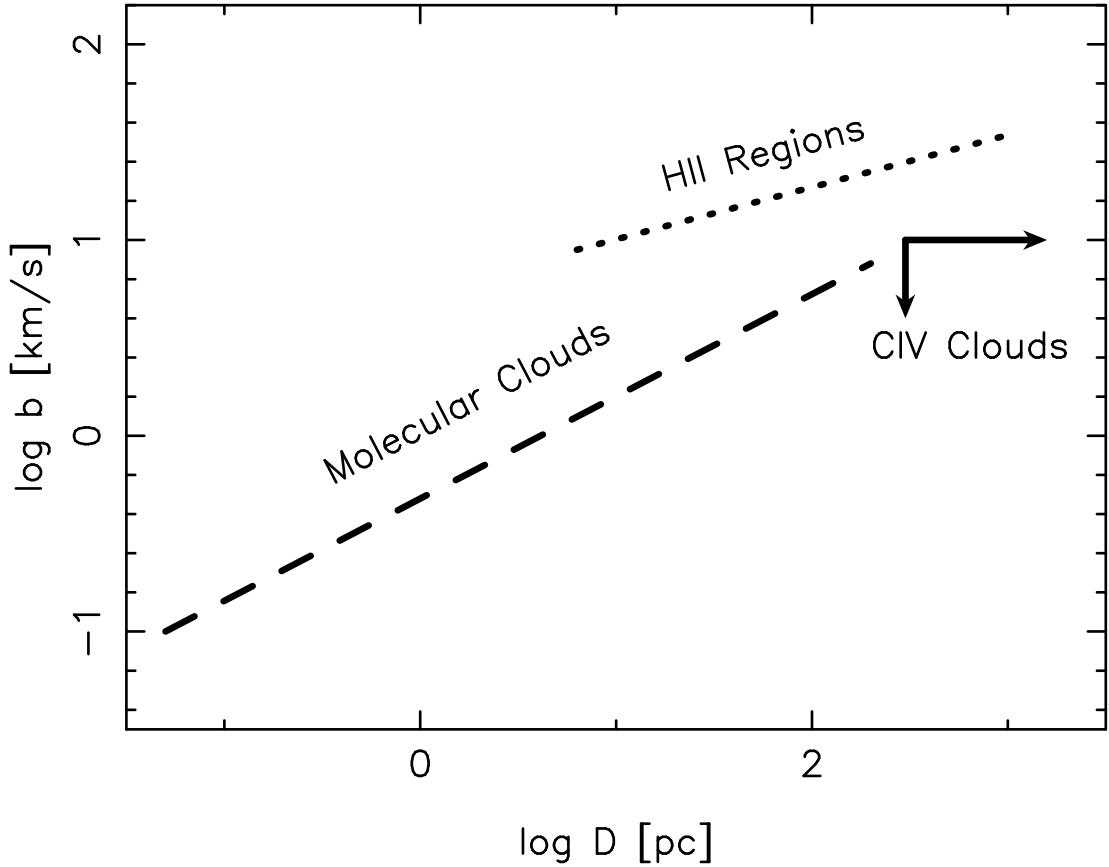}
     \caption{\small The plot shows the relation between the turbulent Doppler 
parameter $b$ and the size $D$ for molecular clouds and HII regions. The solid 
curves schematically represent a compilation of data points from O'Dell 
(1991)). The two arrows corresponds to the upper limits on turbulent line 
widths of individual CIV components and the lower limits on CIV cloud sizes 
corresponding to a 50\% difference in column density. CIV systems are more 
quiescent for their size than HII regions. \label{odellplot}}
 \end{figure}

A comparison of the turbulent Doppler parameters of CIV systems with
other astrophysical environments is shown in Fig.\ref{odellplot}.  The
diagram gives the relation between the turbulent Doppler parameter and
cloud size for a compilation of measurements from molecular clouds
(dashed line) and HII regions (dotted line) by O'Dell (1991). The
figure shows the power law fits to the data given by O'Dell. The arrows
represent the upper limits to the CIV Doppler parameters from Rauch et
al (1996) and the lower limits to the cloud sizes from the present
work. Clearly, the turbulent motion along the line of sight observed
for the C~IV absorbers is  smaller for a given cloud size than the
values found in HII- or star-forming regions.

There is other evidence suggesting that the small scale structure of
the typical high redshift CIV absorbing gas differs from that of
high ionization gas detected locally (i.e., in the Milky Way or the
Magellanic Clouds).  Multiple lines of sight to the Magellanic Clouds
show the gas to be fragmented  on much smaller scales (e.g,  Songaila
et al 1986) than found here; the CIV column density appears to be
fluctuating by 50\% on parsec scales (Wakker et al 1998).  However,
even without invoking evolution in the physical state of the CIV
gas these differences between high and low redshift absorbers are perhaps
not too surprising. We are selecting high z absorption systems
randomly by absorption cross-section  so the observations are weighted
towards large impact parameters from the nearest galaxy.

\subsection{Origin of the velocity structure}

Could the observed velocity scatter be residual turbulence from an earlier
phase of metal ejection ?
In a steady state, the energy transfer rate $\epsilon$ is not only 
the rate at which energy is fed to the gas, but it also equals the 
dissipation rate.
We may estimate the approximate time scale for dissipation, i.e., the
time it takes to transform the mean kinetic energy 1/2 $<v^2>$ in the gas,
at a rate $\epsilon$ into heat,
\begin{eqnarray}
\tau_{\mathrm diss} \sim \frac{1}{2}\frac{<v^2>}{\epsilon} \sim 
9\times10^7{\mathrm years}.
\end{eqnarray}
Apparently, energy is being dissipated at a rate fast enough to destroy
turbulence on a time scale of 100 million years. Thus the turbulence we
observe near the mean redshift $<z>\sim 2.7$ of our sample must have
been produced closer to the epoch of observation and at redshifts lower
than assumed for the Population III process described by, e.g., Gnedin \&
Ostriker (1997). The same authors find that the metallicity of the IGM
produced by a true Pop III star-formation phase (at $z\sim 7-10$) is on the
level of $10^{-4} Z_{\odot}$, two orders of magnitude less than
observed at redshifts $< 4$. Our inferences are {\it consistent} with
most of the metals in the CIV clouds, together with
their turbulent motions, being produced close to the epoch of observation,
at least for $z<5$; this does not preclude the possibility that the 
process started already at much higher redshifts.

An energy input close to the epoch of observation is also indicated by 
what one might call
"cosmic seismometry":  assume that there are recurrent events (SN
explosions, stellar winds, mergers) stirring up the CIV enriched gas,
with a typical interval $\tau_s$ between two such occurrences.  Any
density gradients in the CIV gas caused by such processes would be
damped out by pressure waves propagating with the speed of sound
over a spatial distance  $r$ given by the
product  of the  sound crossing time $\tau_s$ and the sound speed,
$c_s$. Conversely, if there is little structure over a distance $r$,
then there cannot have been a hydrodynamic disturbance during the past
\begin{eqnarray}
\tau_s\sim \frac{r}{c_s} \approx 1.4\times 10^7 
\left(\frac{r}{300\mathrm pc}\right) \left(\frac{c_s}{20 
{\mathrm kms}^{-1}}\right)^{-1} {\mathrm years}.
\end{eqnarray}
This number agrees to within a factor of a few with the dissipation
time scale derived above. We may take these estimates as a measure of the
frequency of disturbing events (probably bursts of star-formation) in the 
vicinity of the absorbing gas. They  indicate that the regions giving rise to 
CIV absorption, while not
{\it currently} experiencing gasdynamical disturbances, may
have done so in the recent past and may continue to do so on a time-scale on 
the order of
10$^7-10^8$ years.

Which physical processes could conceivably  stir up the gas ?  We can
compare the above crude estimates with other galactic time scales
relevant for large scale gasdynamical effects:  

If the turbulent energy in the CIV gas has been injected by galactic outflows
one should expect the duty cycle of star formation to provide the relevant
timescales.
At $10^7$ years the
lifetimes of star-forming molecular clouds (Shu, Adams \& Lizano 1987),
and the dynamical ages for hot superbubbles in local dwarf galaxies
(Martin 1998), show the same order of magnitude as the above measurement.  
Fluctuations in the
star formation rate in nearby spirals (Tomita, Tomita \& Saito 1996;
Hirashita \& Kamaya 2000) and in galactic nuclei (Kr\"ugel \& Tutukov
1992) may reccur every $10^7 - 10^8 $ years, while Glazebrook et
al (1999) and Rocha-Pinto et al (2000) find typical intervals of
$2\times 10^8$ years between maxima in the star-formation rate for their
sample of $z\sim 1$ field galaxies, and the Milky Way, respectively.

As an alternative to stellar feedback, it is conceivable that mergers
or collisions (which Gnedin (1998) requires to account for the observed
metal enrichment) are also capable of stirring the gas to the degree
observed here. The merger rate is highly dependent on the environment
and the redshift. CIV absorbers are thought to reside mostly in or
between low mass halos moving within the filamentary large scale
structure (Rauch, Haehnelt \& Steinmetz 1997). Within the filaments the
average interaction rate may be much higher than in the field.  Menci
\& Valdarnini (1994) predict the mean intervals between merger events
in the more massive filaments to range between $5\times 10^7$ and
$2\times 10^8$ years, depending on the mass of the filamentary
structures, which again is consistent with our estimates.

\section{Conclusions}

Based on the observations of CIV absorption systems in closely spaced
lines of sight to background QSOs we have obtained several new results on the
properties of gas causing the intervening high ionization absorption
systems:

(1) the CIV systems from the combined sample of three lensed
QSOs, UM673, HE1104-1805, and Q1422+231 show that absorption systems 
with redshifts 
between those of the lensing galaxy and the QSO are
increasingly different with redshift. This result shows that the
objects are truly lensed QSOs and not QSO pairs.

(2) We found that the density and velocity structure of CIV absorbing gas at 
high redshift is typically 
featureless  on scales smaller than a few hundred parsecs. Significant 
differences between the
absorption patterns begin to emerge for separations on the order of
kiloparsecs.  

(3) The overall sizes of the CIV systems cannot be probed
very well with lensed QSOs because the beam separations available tend to be
too small, but simple estimates based on the observations presented
here imply that CIV systems extend at least over distances on the order
10 kpc along the line of sight.  This is consistent with other estimates of the
sizes from galaxy-absorber correlations (Bergeron \& Boiss\'e 1991,
Steidel 1993).

(4) A crude measurement of the turbulence in the gas as a function of
projected separation on the plane of the sky indicates a small rate of
energy input, $\epsilon \sim 10^{-3}$ cm$^2$s$^{-3}$. Turbulent
motions along the line of sight as derived from the width of absorption
line profiles also point to a rather quiescent gas, unlike any known
galactic environments.  The CIV absorbing clouds are too 'smooth' for
their size to be, e.g., HII regions.  Both the amplitude of the
turbulence, and the coherence length of the clouds are consistent with
the clouds being affected (produced, stirred, or destroyed) by
star-formation or merger-triggered gas-dynamics on a timescale of
$10^7-10^8$ years.

Realistic numerical studies of the gasdynamics
of these processes on sub-kpc scales, and observations of close QSO images
with high resolution spectrographs and adaptive optics to resolve
images with sub-arcsecond separations would both be desirable for the future. 

\acknowledgments We are grateful to Bob Carswell, John Webb, Andrew
Cooke and Mike Irwin for sharing the VPFIT profile fitting software,
and to Bob Carswell for ample advice and help in using it, and to the
Keck Observatory personnel for their help with the observations.  MR
thanks Matthias Steinmetz for discussions and is grateful to the ESO
Office of Science for travel support, and to the Institute for
Theoretical Physics at the University of Santa Barbara and the
organizers of the workshop on "Galaxy Formation and Evolution" for
hospitality during February and March 2000 (where the research was
supported in part by the National Science Foundation under grant No.
PHY99-07949). The work of WLWS was supported by NSF grant AST-9900733.

\pagebreak


\clearpage
\begin{deluxetable}{rrrrrrrr}
\small
\tablewidth{0pt}
\tablenum{1}
\tablecaption{Comparison of CIV systems towards Q1422+231}
\tablehead{
\colhead{\#} &
\colhead{$N_{A}$} &
\colhead{$N_{B}$} &
\colhead{$\Delta \log N$} &
\colhead{$\overline{v_A} - \overline{v_B}$ } &
\colhead{$\Delta v_A$} &
\colhead{$\Delta v_B$ } &
\colhead{} 
}
\startdata
\multicolumn{8}{c}{ $z= $ {\bf 2.665}  \ \ \ \ \ r = 0.41 h$_{50}^{-1}$ kpc}\\
\hline\hline
 1 &1.06$E+$13& 6.67$E+$12&    0.37 \ \ \  &    16.20 \ \ \  &   150.49&  
223.19& 1 0 0\\
 2 &3.47$E+$12& 4.79$E+$12 &   -0.28 \ \ \  &    -8.72  \ \ \  &    --  &  -- & 
0 0 3\\
\hline
\multicolumn{8}{c}{ $z= $ {\bf 2.68}  \ \ \ \ \ r = 0.395 h$_{50}^{-1}$ kpc}\\
\hline\hline
 1 &1.40$E+$13 &1.81$E+$13 &   -0.22 \ \ \  &    -9.86 \ \ \  &   106.87&   
85.06 & 1 0 0\\
  2& 8.32$E+$13& 1.25$E+$14&   -0.34 \ \ \  &    -0.91  \ \ \  &    -- &  12.36 
& 0 0 3\\
\hline
\multicolumn{8}{c}{ $z= $ {\bf 2.697}  \ \ \ \ \ r = 0.386 h$_{50}^{-1}$ kpc}\\
\hline\hline
 1& 4.04$E+$13& 4.44$E+$13  &  -0.09 \ \ \  &   -11.19\ \ \  &    203.56 & 
212.28 & 1 0 0\\
 2& 1.12$E+$13 &9.08$E+$12  &   0.19  \ \ \  &   -0.28\ \ \  &     10.18 &  
11.63& 0 0 3\\
 3& 2.17$E+$13& 2.45$E+$13 &   -0.11 \ \ \  &    -1.13 \ \ \  &    17.45 &  
38.53& 0 2 0\\
\hline
\multicolumn{8}{c}{ $z= $ {\bf 2.72}  \ \ \ \ \ r = 0.372 h$_{50}^{-1}$ kpc}\\
\hline\hline
 1& 6.69$E+$12& 8.10$E+$12 &    -0.17 \ \ \  &    -5.27 \ \ \  &    29.08 &  
24.72 & 1 0 0\\
\hline
\multicolumn{8}{c}{ $z= $ {\bf 2.75}  \ \ \ \ \ r = 0.356 h$_{50}^{-1}$ kpc}\\
\hline\hline
 1& 5.46$E+$13 &4.57$E+$13   &   0.16 \ \ \  &    -7.50\ \ \  &     81.42 & 
104.69 & 1 0 0\\
 2& 4.11$E+$13 &2.88$E+$13  &    0.30 \ \ \  &     1.55 \ \ \  &    24.72 &  
20.36&  0 2 0\\
 3& 6.76$E+$12& 1.12$E+$13  &  -0.40 \ \ \  &   -12.16 \ \ \  &     --  & 
21.81&  0 0 3\\
\hline\hline
\multicolumn{8}{c}{ $z= $ {\bf 2.77}  \ \ \ \ \ r = 0.342 h$_{50}^{-1}$ kpc}\\
\hline\hline
 1 &1.74$E+$12& 1.41$E+$12 &    0.19 \ \ \  &    -2.91\ \ \  &      --  &  -- & 
1 0 3\\
\hline
\multicolumn{8}{c}{ $z= $ {\bf 2.80}  \ \ \ \ \ r = 0.329 h$_{50}^{-1}$ kpc}\\
\hline\hline
 1& 5.37$E+$12& 7.76$E+$12 &    -0.31 \ \ \  &    21.71 \ \ \  &     -- &  
60.34&  1 0 0\\
\hline
\multicolumn{8}{c}{ $z= $ {\bf 2.90}  \ \ \ \ \ r = 0.277 h$_{50}^{-1}$ kpc}\\
\hline\hline
 1& 1.32$E+$13 &1.41$E+$13    & -0.07\ \ \  &     -0.65\ \ \  &     36.35&   
69.79&  1 0 0\\
\hline
\multicolumn{8}{c}{ $z= $ {\bf 2.95}  \ \ \ \ \ r = 0.252 h$_{50}^{-1}$ kpc}\\
\hline\hline
 1& 5.94$E+$12& 5.60$E+$12 &   0.06 \ \ \  &    -2.68 \ \ \  &   172.30 & 
165.76&  1 0 0\\
 2& 2.63$E+$12& 2.51$E+$12 &    0.05 \ \ \  &    -7.27 \ \ \  &     -- &   -- & 
0 2 3\\
 3& 3.31$E+$12& 3.09$E+$12 &   0.07 \ \ \  &    -0.73 \ \ \  &     -- &   --&  
0 2 3\\
\hline
\multicolumn{8}{c}{ $z= $ {\bf 2.97}  \ \ \ \ \ r = 0.240 h$_{50}^{-1}$ kpc}\\
\hline\hline
 1& 1.14$E+$13& 1.35$E+$13 &  -0.15 \ \ \  &    12.67\ \ \  &    363.50&  
367.14&  1 0 0\\
 2& 3.02$E+$12& 4.14$E+$12 &   -0.27\ \ \  &     -3.20 \ \ \  &     --&   
14.54&  0 2 3\\
 3& 8.40$E+$12& 9.32$E+$12 &  -0.10\ \ \  &     -2.27 \ \ \  &    37.08&   
31.99&  0 2 0\\
 4& 6.31$E+$12& 7.08$E+$12 &   -0.11\ \ \  &     -0.73 \ \ \  &     --&    --&  
0 0 3\\
\hline
\multicolumn{8}{c}{ $z= $ {\bf 3.00}  \ \ \ \ \ r = 0.227 h$_{50}^{-1}$ kpc}\\
\hline\hline
 1& 5.01$E+$12& 5.89$E+$12&    -0.15 \ \ \  &    -2.91 \ \ \  &     --&    --&  
1 0 3\\
\hline
\multicolumn{8}{c}{ $z= $ {\bf 3.06}  \ \ \ \ \ r = 0.199 h$_{50}^{-1}$ kpc}\\
\hline\hline
 1& 1.22$E+$13& 1.82$E+$13&    -0.33 \ \ \  &   -4.80 \ \ \  &   585.23&  
585.23&  1 0 0\\
 2& 9.59$E+$12& 1.47$E+$13&     -0.35 \ \ \  &   -15.26\ \ \  &     91.60&   
82.15&  0 2 0\\
 3& 2.57$E+$12& 3.55$E+$12&    -0.28 \ \ \  &    -8.00 \ \ \  &     --&    --&  
0 2 3\\
 4& 6.78$E+$12& 7.27$E+$12&    -0.07\ \ \  &     -2.30 \ \ \  &    28.35&   
19.63&  0 0 3\\
 5& 2.82$E+$12& 7.41$E+$12&    -0.62\ \ \  &      1.45 \ \ \  &     --&    --&  
0 0 3\\
\hline
\multicolumn{8}{c}{ $z= $ {\bf 3.09}  \ \ \ \ \ r = 0.187 h$_{50}^{-1}$ kpc}\\
\hline\hline
 1& 4.33$E+$13& 4.30$E+$13 &     0.01 \ \ \  &    -4.35 \ \ \  &    87.24&   
87.97&  1 0 0 \\
 2& 9.55$E+$12& 1.05$E+$13&     -0.09 \ \ \  &     0.73 \ \ \  &     -- &   --& 
 0 0 3\\
\hline
\multicolumn{8}{c}{ $z= $ {\bf 3.13}  \ \ \ \ \ r = 0.169 h$_{50}^{-1}$ kpc}\\
\hline\hline
 1& 3.38$E+$13& 2.95$E+$13 &   0.13 \ \ \  &    -9.00 \ \ \  &   402.76&  
423.84&  1 0 0\\
 2& 2.36$E+$13& 2.09$E+$13 &     0.12 \ \ \  &   -14.43 \ \ \  &   148.31&  
146.85&  0 2 0\\
 3 &1.02$E+$13 &8.67$E+$12&    0.15  \ \ \  &   -2.83 \ \ \  &   103.96&   
87.97& 0 2 0\\
 4& 6.92$E+$12& 7.29$E+$12 &   -0.05 \ \ \  &    -1.26 \ \ \  &     --&   17.45 
& 0 0 3\\
 5& 4.68$E+$12& 5.25$E+$12&    -0.11 \ \ \  &    -2.18  \ \ \  &    --&    --&  
0 0 3\\
\hline
\multicolumn{8}{c}{ $z= $ {\bf 3.27}  \ \ \ \ \ r = 0.117 h$_{50}^{-1}$ kpc}\\
\hline\hline
 1& 4.68$E+$12& 4.47$E+$12 &    0.05  \ \ \  &    -- \ \ \  &     -- &   -- & 1 
0 3\\
\hline
\multicolumn{8}{c}{ $z= $ {\bf 3.41}  \ \ \ \ \ r = 0.066 h$_{50}^{-1}$ kpc}\\
\hline\hline
 1& 8.13$E+$12& 8.48$E+$12  &   -0.04 \ \ \  &    -6.35 \ \ \  &     --&    
5.09  &1 0 0\\
\hline
\multicolumn{8}{c}{ $z= $ {\bf 3.53}  \ \ \ \ \ r = 0.027 h$_{50}^{-1}$ kpc}\\
\hline\hline
 1& 2.30$E+$14& 2.38$E+$14&    -0.03 \ \ \  &     2.81\ \ \  &    426.02&  
434.02&  1 0 0\\
 2& 4.53$E+$13& 4.55$E+$13&     0.00 \ \ \  &    -2.18\ \ \  &     13.81&   
11.63& 0 0 3\\
 3 &2.41$E+$13& 2.73$E+$13 &   -0.12 \ \ \  &    -2.63 \ \ \  &    15.27&   
30.53&  0 0 3\\
 4 &6.26$E+$13& 6.88$E+$13&    -0.09 \ \ \  &    -1.30 \ \ \  &     9.45&    
8.72&  0 0 3\\
 5 &.3.84$E+$13& 3.77$E+$13 &    0.02 \ \ \  &    -1.78 \ \ \  &     9.45 &   
8.00 & 0 0 3\\
\hline
\multicolumn{8}{c}{ $z= $ {\bf 3.58}  \ \ \ \ \ r = 0.016 h$_{50}^{-1}$ kpc}\\
\hline\hline
 1& 2.25$E+$13& 2.31$E+$13&     -0.03 \ \ \  &     6.12 \ \ \  &   145.40&  
148.31&  1 0 0\\
 2& 1.55$E+$13& 1.70$E+$13 &    -0.09 \ \ \  &    -0.73 \ \ \  &    --&    --&  
0 0 3\\
 3& 6.98$E+$12& 6.08$E+$12&      0.13 \ \ \  &     0.26 \ \ \  &     6.54&   
12.36&  0 0 3\\
\hline
\multicolumn{8}{c}{ $z= $ {\bf 3.62}  \ \ \ \ \ r = 0.001 h$_{50}^{-1}$ kpc}\\
\hline\hline
 1& 1.01$E+$14& 9.43$E+$13&     0.07 \ \ \  &    -2.92\ \ \  &    74.88&   
58.16 & 1 0 0\\
 2& 339$E+$13& 2.95$E+$13&     0.13 \ \ \  &    -1.13\ \ \  &      3.63&    -- 
& 0 0 3\\
 3& 5.41$E+$13& 5.94$E+$13&    -0.09 \ \ \  &    -1.61\ \ \  &      0.73&   
18.90 & 0 0 3\\
\enddata
\end{deluxetable} 

\clearpage

\clearpage
\begin{deluxetable}{rrrrrrrr}
\small
\tablewidth{0pt}
\tablenum{2}
\tablecaption{Comparison of CIV systems towards HE1104--1805}
\tablehead{
\colhead{\#} &
\colhead{$N_{A}$} &
\colhead{$N_{B}$} &
\colhead{$\Delta \log N$} &
\colhead{$\overline{v_A} - \overline{v_B}$ } &
\colhead{$\Delta v_A$} &
\colhead{$\Delta v_B$ } &
\colhead{} 
}
\startdata
\multicolumn{8}{c}{ $z= $ {\bf 1.661}  \ \ \ \ \ r = 4.7 h$_{50}^{-1}$ kpc}\\
\hline\hline
 1 &2.75$E+$14&4.22$E+$14&     -0.35\ \ \  &   62.66\ \ \   &  469.64&  420.21& 
 1 0 0\\
 2 &1.59$E+$14&3.54$E+$14&    -0.55\ \ \   &  -24.29\ \ \   &  259.54&  207.20& 
 0 0 0\\
 3 &1.16$E+$14&6.73$E+$13&     0.42\ \ \  &  -20.23\ \ \ &  131.59&   61.79&  0 
2 0\\
 4 &1.21$E+$14&3.00$E+$14&    -0.60\ \ \ &  -44.14\ \ \ &   40.71&   59.61&  0 
2 0\\
 5 &7.76$E+$13&2.59$E+$13&    0.67\ \ \ &  -10.80\ \ \ &    -- \ \ \ &   
13.81&  0 2 0\\
 6 &1.05$E+$14&5.89$E+$13&    0.44\ \ \ &  -27.63\ \ \ &    -- \ \ \ &    -- \ 
\ \ &  0 0 3\\
\hline
\multicolumn{8}{c}{ $z= $ {\bf 1.747}  \ \ \ \ \ r = 3.9 h$_{50}^{-1}$ kpc}\\
\hline\hline
 1 &8.32$E+$12&4.68$E+$12&      0.44\ \ \ &   -1.45\ \ \ &    -- \ \ \ &    -- 
\ \ \ &  1 0 3\\
\hline
\multicolumn{8}{c}{ $z= ${\bf  1.859}  \ \ \ \ \ r = 2.9 h$_{50}^{-1}$ kpc}\\
\hline\hline
 1 &4.45$E+$13&8.25$E+$13&      -0.46\ \ \ &   50.37\ \ \ &  251.54&  303.16&  
1 0 0\\
 2 &3.54$E+$13&7.23$E+$13&      -0.51\ \ \ &   38.52\ \ \ &   76.33&  179.57&  
0 2 0\\
 3 &9.12$E+$12&1.02$E+$13&      -0.10\ \ \ &    6.37\ \ \ &    -- \ \ \ &   
33.44&  0 2 0\\
 4 &9.12$E+$12&9.33$E+$12&      -0.02\ \ \ &    3.63\ \ \ &    -- \ \ \ &    -- 
\ \ \ &  0 0 3\\
\hline
\multicolumn{8}{c}{ $z= $ {\bf 2.01}  \ \ \ \ \ r = 1.76 h$_{50}^{-1}$ kpc}\\
\hline\hline
 1 &not det.&1.32$E+$13&     -1.00\ \ \ &  -\ \ \ &  - &  83.61&  1 0 0\\
\hline
\multicolumn{8}{c}{ $z= $ {\bf 2.05}  \ \ \ \ \ r = 1.48 h$_{50}^{-1}$ kpc}\\
\hline\hline
 1 &not det.&7.10$E+$13&     -1.00\ \ \ &  -\ \ \ &  - &  26.17&  1 0 0\\
\hline
\multicolumn{8}{c}{ $z= $ {\bf 2.20}  \ \ \ \ \ r = 0.59 h$_{50}^{-1}$ kpc}\\
\hline\hline
 1 &4.47$E+$13&9.13$E+$13&     -0.51\ \ \ &  -16.96\ \ \ &  175.21&  210.83&  1 
0 0\\
 2 &2.22$E+$13&7.19$E+$13&     -0.69\ \ \ &   -7.57\ \ \ &   29.81&    8.72&  0 
2 0\\
 3 &3.22$E+$12&1.58$E+$12&      0.51\ \ \ &  -10.83\ \ \ &    7.27&    -- \ \ \ 
&  0 2 0\\
\hline
\multicolumn{8}{c}{ $z= ${\bf  2.298}  \ \ \ \ \ r = 0.06 h$_{50}^{-1}$ kpc}\\
\hline\hline
 1 &9.84$E+$13&1.41$E+$14&     -0.30\ \ \ &    9.16\ \ \ &  300.25&  330.79&  1 
0 0\\
 2 &7.43$E+$13&1.15$E+$14&     -0.35\ \ \ &   -0.17\ \ \ &  109.78&  113.41&  0 
2 0\\
 3 &2.41$E+$13&2.62$E+$13&     -0.08\ \ \ &   -4.30\ \ \ &  141.76&  173.03&  0 
2 0\\
 4 &4.37$E+$13&3.39$E+$13&      0.22\ \ \ &    6.54\ \ \ &    -- \ \ \ &    -- 
\ \ \ &  0 0 3\\
 5 &2.56$E+$13&7.02$E+$13&     -0.64\ \ \ &   10.79\ \ \ &   10.18&   12.36&  0 
0 3\\
 6 &1.45$E+$12&6.46$E+$11&      0.55\ \ \ &   -4.36\ \ \ &    -- \ \ \ &    -- 
\ \ \ &  0 0 3\\
 7 &1.50$E+$13&9.77$E+$12&      0.35\ \ \ &   -0.26\ \ \ &    5.82&    -- \ \ \ 
&  0 0 3\\
 8 &4.17$E+$12&1.13$E+$13&     -0.63\ \ \ &    5.90\ \ \ &    -- \ \ \ &    
8.00&  0 0 3\\
 9 &1.74$E+$12&1.01$E+$13&     -0.83\ \ \ &    -0.17\ \ \ &    -- \ \ \ &    0.73 
\ \ \ &  0 0 3\\
\hline
\multicolumn{8}{c}{ $z= $ {\bf 2.314}  \ \ \ \ \ \ \ \ \ $z_{\mathrm abs}>z_{em}$}\\
\hline\hline
 1 &7.58$E+$12 &7.72$E+$12&    -0.02\ \ \ &    1.61\ \ \ &   43.62&   50.89&  1 
0 0\\
 2 &6.46$E+$12 &6.76$E+$12&    -0.05\ \ \ &    1.45\ \ \ &    -- \ \ \ &    -- 
\ \ \ &  0 0 3\\
 3 &1.12$E+$12 &9.55$E+$11&     0.15\ \ \ &   -5.82\ \ \ &    -- \ \ \ &    -- 
\ \ \ &  0 0 3\\
\enddata
\end{deluxetable} 

\clearpage

\begin{deluxetable}{rrrrrrrr}
\small
\tablewidth{0pt}
\tablenum{3}
\tablecaption{Comparison of CIV systems towards UM 673 A and B}
\tablehead{
\colhead{\#} &
\colhead{$N_{A}$} &
\colhead{$N_{B}$} &
\colhead{$\Delta \log N$} &
\colhead{$\overline{v_A} - \overline{v_B}$ } &
\colhead{$\Delta v_A$} &
\colhead{$\Delta v_B$ } &
\colhead{} 
}
\startdata
\multicolumn{8}{c}{ $z= ${\bf  1.940}  \ \ \ \ \ r = 1.70 h$_{50}^{-1}$ kpc}\\
\hline\hline
 1 &5.41$E+$14 &4.62$E+$14&    0.15\ \ \ &   62.51\ \ \ &  471.10&  486.36&  1 
0 0\\
 2 &3.41$E+$14 &3.87$E+$14&    -0.12\ \ \ &   -8.80\ \ \ &  178.84&  210.83&  0 
2 0\\
 3 &2.07$E+$14 &3.27$E+$14&    -0.37\ \ \ &  -26.94\ \ \ &   85.06&   81.42&  0 
0 0\\
 4 &3.47$E+$13 &2.04$E+$13&     0.41\ \ \ &   -7.27\ \ \ &    -- \ \ \ &    -- 
\ \ \ &  0 0 3\\
 5 &2.82$E+$13 &2.57$E+$13&     0.09\ \ \ &  -21.08\ \ \ &    -- \ \ \ &    -- 
\ \ \ &  0 0 3\\
 6 &1.99$E+$14 &7.44$E+$13&     0.63\ \ \ &   -5.93\ \ \ &  117.77&  117.77&  0 
2 0\\
 7 &7.08$E+$12 &9.55$E+$12&    -0.26\ \ \ &   -4.36\ \ \ &    -- \ \ \ &    -- 
\ \ \ &  0 0 3\\
 8 &1.05$E+$14 &4.79$E+$13&     0.54\ \ \ &   -2.91\ \ \ &    -- \ \ \ &    -- 
\ \ \ &  0 0 3\\
 9 &2.19$E+$13 &1.15$E+$13&     0.48\ \ \ &   10.90\ \ \ &    -- \ \ \ &    -- 
\ \ \ &  0 0 3\\
\hline
\multicolumn{8}{c}{ $z= $ {\bf 2.060}  \ \ \ \ \ r = 1.34 h$_{50}^{-1}$ kpc}\\
\hline\hline
 1 &3.80$E+$13 &5.26$E+$13&     -0.28\ \ \ &  -68.66\ \ \ &  605.59&  615.04&  1 
0 0\\
 2 &1.25$E+$13 &1.12$E+$13&     0.10\ \ \ &    4.74\ \ \ &   28.35&    -- \ \ \ 
&  0 2 0\\
 3 &1.20$E+$13 &1.12$E+$13&     0.07\ \ \ &    3.63\ \ \ &   -- \ \ \ &   -- \ 
\ \ &  0 0 3\\
 4 &2.55$E+$13 &4.13$E+$13&    -0.38\ \ \ &   -5.82\ \ \ &   51.62&   57.43&  0 
2 0\\
 5 &1.23$E+$13 &2.24$E+$13&    -0.45\ \ \ &  -19.63\ \ \ &   -- \ \ \ &   -- \ 
\ \ &  0 0 3\\
 6 &1.45$E+$12 &3.47$E+$12&    -0.58\ \ \ &   -5.82\ \ \ &   -- \ \ \ &   -- \ 
\ \ &  0 0 3\\
\hline
\multicolumn{8}{c}{ $z= ${\bf  2.355}  \ \ \ \ \ r = 0.63 h$_{50}^{-1}$ kpc}\\
\hline\hline
 1 &3.79$E+$14 &3.79$E+$14&     0.00\ \ \ &  -17.32\ \ \ &  285.71&  311.88&  1 
0 0\\
 2 &3.72$E+$14 &3.69$E+$14&     0.01\ \ \ &  -15.15\ \ \ &  205.74&  183.93&  0 
2 0\\
 3 &6.17$E+$12 &5.37$E+$12&     0.13\ \ \ &    3.63\ \ \ &   -- \ \ \ &   -- \ 
\ \ &  0 2 3\\
\hline
\multicolumn{8}{c}{ $z= ${\bf  2.665} \ \ \ \ \ r =  0.08 h$_{50}^{-1}$ kpc}\\
\hline\hline
 1 &1.30$E+$13 &1.32$E+$13&    -0.02\ \ \ &    2.24\ \ \ &   49.44&   13.09&  1 
2 0\\
 2 &1.15$E+$13 &1.32$E+$13&    -0.13\ \ \ &   -3.40\ \ \ &   -- \ \ \ &   
13.09&  0 0 0\\
 3 &1.15$E+$13 &1.05$E+$13&     0.09\ \ \ &   -0.73\ \ \ &   -- \ \ \ &   -- \ 
\ \ &  0 0 3\\
\hline
\multicolumn{8}{c}{ $z= ${\bf  2.71}  \ \ \ \ \ \ \ \ \ r =  0.02  
h$_{50}^{-1}$ kpc}\\
\hline\hline
 1 &9.25$E+$12 &9.85$E+$12&    -0.06\ \ \ &    0.61\ \ \ &    8.72&   15.99&  1 
2 0\\
\hline
\multicolumn{8}{c}{ $z= $ {\bf 2.736}  \ \ \ \ \ \ \ \ \ $z_{\mathrm abs}>z_{em}$}\\
\hline\hline
 1 &2.39$E+$13 &2.06$E+$13&     0.13\ \ \ &   18.72\ \ \ &  272.62&  305.34&  1 
0 0\\
 2 &1.53$E+$13 &1.52$E+$13&     0.01\ \ \ &    0.08\ \ \ &   16.72&   18.90&  0 
2 0\\
 3 &8.71$E+$12 &8.32$E+$12&     0.05\ \ \ &    1.45\ \ \ &   -- \ \ \ &   -- \ 
\ \ &  0 0 3\\
 4 &8.54$E+$12 &5.40$E+$12&     0.37\ \ \ &  -19.97\ \ \ &   36.35&   69.07&  0 
2 0\\
 5 &4.07$E+$12 &2.24$E+$12&     0.45\ \ \ &    1.45\ \ \ &   -- \ \ \ &   -- \ 
\ \ &  0 0 3\\
 6 &4.47$E+$12 &3.16$E+$12&     0.29\ \ \ &  -31.26\ \ \ &   -- \ \ \ &   -- \ 
\ \ &  0 0 3\\
\enddata
\end{deluxetable} 

\clearpage

\section{figures}

\begin{figure}
    \centering
  \includegraphics[width=14.cm, angle=0]{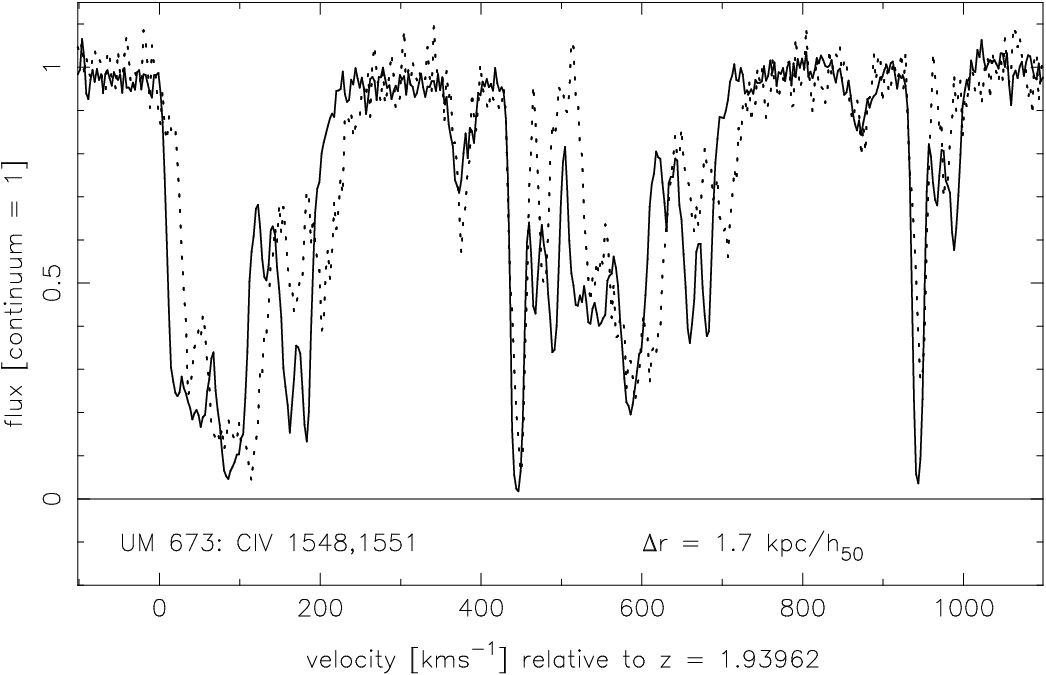}
  \includegraphics[width=14.cm,angle=0]{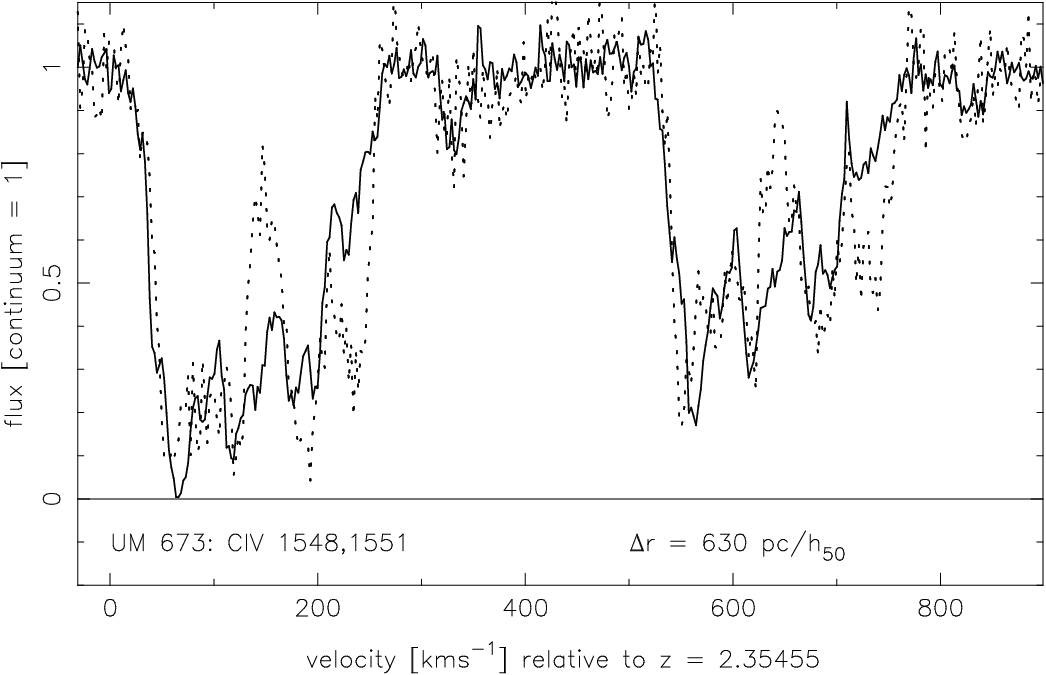}
         \caption{\small Two examples of strong CIV systems at redshifts 1.940 (top; (separation
between the lines of sight: 1.7 $h_{50}^{-1}$ kpc) and 2.355 (bottom; (separation
between the lines of sight: 0.63 $h_{50}^{-1}$ kpc) towards UM673. Solid line and
dotted line show the spectra of the A and B image, respectively. The
velocity scale applies to the rest frame of the systems.  In the upper
image note the different
amounts of velocity shear ($v_A -v_B$) for the absorption clump with
$0<v<250$kms$^{-1}$ and the sharp, nearly unshifted absorption line near 450 
kms$^{-1}$. This absorption signature could be caused by a merger of two 
clumps of gas with
different internal kinematics.
}
\end{figure} 

\end{document}